\documentclass[12pt]{article}
\usepackage{latexsym}
\usepackage{amssymb}
\usepackage{amsmath}
\usepackage{amsfonts}

\numberwithin{equation}{section}

\newcommand{\be}{\begin{equation}}
\newcommand{\ee}{\end{equation}}

\newcommand{\Tr}{{\rm Tr}}

\def\cN{{\cal N}}
\def\bea{\begin{eqnarray}}
\def\eea{\end{eqnarray}}
\def\nn{\nonumber}

\begin{document}

\begin{titlepage}

\vspace{1cm}

\begin{center}
{\Large\bf Two-loop effective potentials in general
\\[3mm]
$\cN=2$, $d=3$ chiral superfield model} \vspace{1.5cm}
 \\[0.5cm]
 {\bf
 I.L. Buchbinder $^+$\footnote{joseph@tspu.edu.ru},
 B.S. Merzlikin $^{+\dag}$\footnote{merzlikin@mph.phtd.tpu.ru},
 I.B. Samsonov $^{*}$\footnote{samsonov@mph.phtd.tpu.ru, on leave from
Tomsk Polytechnic University, 634050 Tomsk, Russia.}}
\\[3mm]
 {\it $^+$ Department of Theoretical Physics, Tomsk State Pedagogical
 University,\\ Tomsk 634061, Russia\\[2mm]
 $^\dag$
 Department of Higher Mathematics and Mathematical Physics,\\
\it Tomsk Polytechnic University, 634050 Tomsk, Russia
 \\[2mm]
 $^*$ INFN, Sezione di Padova, via F. Marzolo 8, 35131 Padova, Italy}
 \\[0.8cm]
\bf Abstract
\end{center}
We study local superspace contributions to low-energy effective action
in general chiral three-dimensional superfield model. The effective K\"ahler and
chiral potentials are computed in explicit form up to the two-loop order.
In accordance with the non-renormalization theorem, the
ultraviolet divergences appear only in the full superspace while
the effective chiral potential receives only finite quantum
contributions in the massless case. As an application,
the two-loop effective scalar potential
is found for the three-dimensional ${\cal N}=2$ supersymmetric
Wess-Zumino model.
\end{titlepage}
\setcounter{footnote}{0}

\section{Introduction}
\setcounter{equation}{0}
The interest to three-dimensional supersymmetric field models is
partly motivated by recent achievements in constructing field
theories modelling multiple M2 branes (see, e.g., \cite{BLG1,BLG2,BLG3,ABJM1}). These
are three-dimensional superconformal field theories of
Chern-Simons gauge fields interacting with matter
in a special way such that superconformal invariance
and $\cN=6$ (or even $\cN=8$) extended supersymmetry are respected.
There is very significant progress in studying correlation functions and
scattering amplitudes in the ABJM-like models (see, e.g., recent papers
\cite{Penati1,Penati2,Penati3,Minahan1}),
but the problem of low-energy effective action in such theories is
still purely understood. One of the subtle questions in defining the low-energy
effective action in the ABJM theory is how to separate the light and heavy
degrees of freedom when the scalar fields acquire some vacuum.
Indeed, in \cite{M2D2-1}
it was shown that the Higgs mechanism for the
ABJM-like models works differently from the standard case of
SYM-matter models. Once the scalar fields acquire some vevs, this
novel Higgs mechanism turns the ABJM action into the $\cN=8$ SYM
model with higher derivative corrections. So, one is forced
to study the effective action in the three-dimensional extended SYM models
\cite{BPS1,BPS2} rather in the ABJM model itself.

The Chern-Simons fields in the ABJM action make the M2 brane
description very elegant because the supersymmetries, R-symmetry
and conformal invariance become quite explicit \cite{Schwarz1,Schwarz2}. However,
these fields are non-dynamical and can be, in principle,
eliminated by fixing the gauge symmetry. Once the gauge symmetry
is fixed, one is left with a three-dimensional non-linear
supersymmetric sigma-model in which the symmetries of the M2
brane become non-manifest. Quantum aspects of such a
sigma-model (and, in particular, low-energy effective action) can
be investigated by standard methods of quantum field theory.

Keeping these motivations in mind, we initiate the studies of
low-energy effective action in $\cN=2$, $d=3$ supersymmetric
sigma-models. In the present paper we restrict ourself to the
model with one chiral superfield $\Phi$ which is described by
the following general action
\be
S=-\int d^3xd^4\theta\,K(\Phi,\bar\Phi)
-\left(\int d^3x d^2\theta\, W(\Phi)  +c.c.\right)\,.
\label{class}
\ee
Here $K(\Phi,\bar\Phi)$ is the K\"ahler
potential and $W(\Phi)$ is chiral potential. The sigma-model case
corresponds to $W\equiv 0$, but we keep it non-vanishing for the
sake of generality. Another particular case with
\be
K_{\rm WZ}=\Phi\bar\Phi\,,\qquad
W_{\rm WZ}=\frac m2 \Phi^2+\frac\lambda 6\Phi^3
+\frac g{24}\Phi^4\,,
\label{WZ}
\ee
corresponds to the classical action of $\cN=2$, $d=3$ Wess-Zumino
model.

The aim of this paper is to study some aspects of superfield quantum
effective action in the model under consideration. In general, the
effective action is extremely complicated non-local functional of
background superfields. To be more precise, we are interested in the
local part of the quantum effective action in the model
(\ref{class}) which is described by the effective K\"ahler and
chiral potentials,
\be \Gamma=-\int d^3x d^4\theta\, K_{\rm
eff}(\Phi,\bar\Phi) -\left(\int d^3x d^2\theta\, W_{\rm
eff}(\Phi)+c.c.\right)\,.\ee
All non-local terms or the terms with
derivatives of background superfields are out of our
approximation.\footnote{It means that all possible derivative
dependent or space-time non-local contributions to effective action
are systematically neglected.} In the present paper we compute both
one- and two-loop quantum corrections to $K_{\rm eff}$ and $W_{\rm
eff}$ and briefly discuss the application of these results to the
three-dimensional Wess-Zumino model. In particular, effective scalar
potential in the Wess-Zumino models is found.

The effective K\"ahler and chiral potentials for the
four-dimensional analog of the model (\ref{class}) were studied in
\cite{BKY,PW,West1,West2,BKP1,BKP2,BP1,BP2,BCP1,BCP2,Brig,Nibbelink}.
In these papers it
was shown that the quantum divergences appear only in the sector
of the effective K\"ahler potential while the effective chiral
potential is finite. These results are in complete agreement with
the non-renormalization theorem \cite{GGRS,book} which applies for the
three-dimensional models as well. In the present work, for the
three-dimensional general chiral superfield model we show that
UV divergences appear only in the Feynman graphs contributing
to the effective K\"ahler potential while the effective chiral potential
UV finite. Such finite quantum corrections in the
chiral sector arise only in the massless case.

We point out that the effective K\"ahler potential in the $\cN=1$,
$d=3$ scalar superfield model was studied in \cite{Lehum1,Lehum2,Lehum3}.
In the present work we extend some of the results of
these papers to the $\cN=2$ supersymmetric case.

The rest of this paper is organized as follows. In Sect.\ 2 we
find one- and two-loop quantum contributions to the effective K\"ahler
potential while similar computations for the effective chiral
potential are given in Sect.\ 3. In Sect.\ 4 we apply these
general results for obtaining effective scalar
potential in the $\cN=2$, $d=3$ Wess-Zumino models. The details of
computations of Feynman graphs
contributing to the effective K\"ahler and chiral
potentials are given in the Appendices.

Throughout this work we employ the $\cN=2$, $d=3$ superspace
notations from \cite{BPS1,BPS2}.

\section{Effective K\"ahler potential}
\setcounter{equation}{0}
Within the loop expansion of the effective action
the effective K\"ahler potential can be represented by a series
\be
K_{\rm eff}=K+K^{(1)}+K^{(2)}+\ldots\,,
\ee
where $K$ is the classical K\"ahler potential and $K^{(1)}$, $K^{(2)}$ correspond to
one- and two-loop quantum contributions. The dots stand for higher
loops. In the present paper we restrict ourself to the two-loop
approximation.
\subsection{One-loop contributions}
Let us split the superfield $\Phi$ into the ``background'' $\Phi$ and
``quantum'' $\phi$ parts, $\Phi\to\Phi+\phi$.
For computing the one-loop effective action it is sufficient to
expand the classical action (\ref{class}) up to the second order
in quantum superfields,
\be
S=-\int d^3xd^4\theta[\frac12 \phi^2 K''_{\Phi\Phi}
+\frac12 \bar \phi^2 K''_{\bar\Phi\bar\Phi}
+\phi\bar\phi K''_{\Phi\bar\Phi}]
-\left(\int d^3x d^2\theta\, \frac12\phi^2 W''(\Phi)  +c.c.\right)
+\ldots\,,
\label{class2}
\ee
where dots stand for higher order terms with respect to the
quantum fields. For computing $K_{\rm
eff}(\Phi,\bar\Phi)$ it is sufficient to consider {\it constant}
background fields,
\be
D_\alpha\Phi=0\,,\qquad
\bar D_\alpha \bar\Phi=0\,.
\label{const-back}
\ee
In this case the terms involving $K''_{\Phi\Phi}$ and
$K''_{\bar\Phi\bar\Phi}$ in (\ref{class2}) vanish.

The one-loop effective action is given by
\be
\Gamma^{(1)}=\frac i2\Tr\ln H\,,
\label{G1}
\ee
where $H$ is an operator which defines the part of the action
which is quadratic with respect to the background superfields,
\be
H=\left(
\begin{array}{cc}
-W'' & \frac14K''_{\Phi\bar\Phi}\bar D^2 \\
\frac14K''_{\Phi\bar\Phi} D^2 & \bar W'' \\
\end{array}
\right)\,.
\ee
We represent this operator as a sum of two terms,
\be
H=H_0+H_1\,,\qquad
H_0=\left(
\begin{array}{cc}
0 & \frac14K''_{\Phi\bar\Phi}\bar D^2 \\
\frac14K''_{\Phi\bar\Phi} D^2 & 0 \\
\end{array}
\right)\,,\qquad
H_1=\left(
\begin{array}{cc}
-W'' & 0 \\
0 & \bar W'' \\
\end{array}
\right)\,,
\ee
and expand the logarithm in (\ref{G1}),
\be
\Gamma^{(1)}=\frac i2\Tr\ln H_0
+\frac i2\Tr\sum_{n=1}^\infty\frac{(-1)^{n+1}}{n}
(H_0^{-1}H_1)^n\,.
\label{8}
\ee
For the constant background fields, the first term in the rhs of
(\ref8) reads
\be
\frac i2\Tr\ln H_0 =\frac i4 \Tr \ln H_0^2
=\frac i4 \Tr_+ \ln((K''_{\Phi\bar\Phi})^2\square)+c.c.
\label{9}
\ee
Here $\Tr_+$ denotes the functional trace over the chiral
superfields. The box operator in the rhs in (\ref{9})
originates from the identity
\be
\frac1{16}\bar D^2 D^2 \Phi=\square \Phi\,,
\label{box}
\ee
which holds for any chiral superfield $\Phi$ and follows from
the anticommutation relations of the Grassmann derivatives,
$\{D_\alpha,\bar D_\beta \}=-2i\partial_{\alpha\beta}$.

Using the inverse operator
\be
H_0^{-1}=\left(
\begin{array}{cc}
0 & \frac14\frac{\bar D^2}{K''_{\Phi\bar\Phi}\square} \\
\frac14\frac{ D^2}{K''_{\Phi\bar\Phi}\square} &0
\end{array}
\right),
\ee
for the second term in the rhs of (\ref{8}) we get
\bea
&&\frac i2\Tr\sum_{n=1}^\infty\frac{(-1)^{n+1}}n
\left(
\begin{array}{cc}
0 & \frac14\frac{\bar W'' \bar D^2}{K''_{\Phi\bar\Phi}\square} \\
\frac14\frac{ W'' D^2}{K''_{\Phi\bar\Phi}\square} &0
\end{array}
\right)^n
=\frac i4\Tr\sum_{n=1}^\infty\frac{(-1)^{n+1}}n
\left(
\begin{array}{cc}
\left|\frac{W''}{K''_{\Phi\bar\Phi}}\right|^2\frac1\square & 0\\
0 & \left|\frac{W''}{K''_{\Phi\bar\Phi}}\right|^2\frac1\square
\end{array}
\right)^n\nn\\&=&\frac
i4\Tr_+\ln\left(1+\left|\frac{W''}{K''_{\Phi\bar\Phi}}\right|^2\frac1\square\right)
+c.c.
\label{11}
\eea
Combining (\ref9) and (\ref{11}) together, we get the following formal
expression for the one-loop effective action
\be
\Gamma^{(1)}=\frac i4\Tr_+\ln(\square+M^2) +c.c.\,,
\label{12}
\ee
where the effective mass squared is
\be
M^2=\left|\frac{W''}{K''_{\Phi\bar\Phi}}\right|^2\,.
\ee

The functional trace over chiral superfields in (\ref{12}) can be
written explicitly as
\be
\Tr_+\ln(\square+M^2)=\int d^5 z_1 d^5 z_2\,
\delta_+(z_2,z_1) \ln(\square+M^2) \delta_+(z_1,z_2)\,.
\ee
To compute this expression we apply the following identity for the
chiral delta-function,
\be
\delta_+(z_2,z_1)=\frac1{16}\frac{\bar D^2
D^2}{\square}\delta_+(z_2,z_1)\,,\qquad
\delta_+(z_1,z_2)=-\frac14\bar D^2 \delta^7(z_1-z_2)\,.
\label{chiral-delta}
\ee
Then, we restore full superspace measure using the extra $\bar
D^2$-operators,
\be
d^7z=-\frac14 \bar D^2 d^5z\,,
\label{measure}
\ee
 and integrate by
parts the $D^2$-operator,
\bea
\Tr_+\ln(\square+M^2) &=& \int d^5 z_1 d^5 z_2
(-\frac14\bar D^2_{(2)})(-\frac14\bar D^2_{(1)})(-\frac14 D^2_{(1)})
\\&&\times
\frac1{\square}\delta^7(z_2-z_1) \ln(\square+M^2) \delta_+(z_1,z_2)
\nn\\&=& \int d^7 z_1 d^7 z_2(-\frac14 D^2_{(1)})\frac1{\square}
\delta^7(z_2-z_1) \ln(\square+M^2) \delta_+(z_1,z_2)
\nn\\&=&\int d^7 z_1 d^7 z_2\,
\delta^7(z_2-z_1) \ln(\square+M^2)(-\frac14 D^2_{(1)})(-\frac14\bar D^2_{(1)})\frac1{\square} \delta^7(z_1-z_2)
\nn
\,.
\eea
Now, using standard identity
\be
\delta^4(\theta_1-\theta_2)\frac1{16}D^2 \bar D^2
\delta^7(z_1-z_2)=\delta^7(z_1-z_2)\,,
\label{id}
\ee
we integrate over one set of Grassmann variables,
\be
\Tr_+\ln(\square+M^2)=\int d^3x_1 d^3x_2 d^4\theta\,
\delta^3(x_1-x_2)\ln(\square+M^2)\frac1\square\delta^3(x_2-x_1)\,.
\ee
Passing to the momentum space, we compute the momentum integral,
\be
-\int \frac{d^3p}{(2\pi)^3}\frac1{p^2}\ln\left(1-\frac{M^2}{p^2}\right)
=\frac i{2\pi}|M|\,.
\ee

As a result, we get the following answer for the one-loop
K\"ahler potential,
\be
K^{(1)}(\Phi,\bar\Phi)=\frac1{4\pi}
|M|=\frac1{4\pi}\frac{|W''|}{K''_{\Phi\bar\Phi}}\,.
\label{K}
\ee
We point out that the one-loop effective action is finite because only
two-loop UV quantum divergences can appear in three-dimensional
field theories. Hence, (\ref{K}) is a finite renormalization of
the classical K\"ahler potential in (\ref{class}).

\subsection{Two-loop contributions}
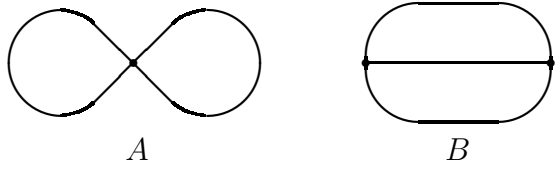
\begin{figure}[tb]
\begin{center}
\begin{picture}(260,70)
\thicklines
\put(40,40){\oval(40,40)[l]}
\put(95,40){\oval(40,40)[r]}
\qbezier(40,20)(48,21)(52,25)
\qbezier(40,60)(48,59)(52,55)
\put(52,25){\line(1,1){30}}
\put(52,55){\line(1,-1){30}}
\qbezier(94,20)(86,21)(82,25)
\qbezier(94,60)(86,59)(82,55)
\put(67,40){\circle*{3}}
\put(64,3){$A$}
\put(190,40){\oval(70,45)}
\put(155,40){\line(1,0){70}}
\put(155,40){\circle*{3}}
\put(225,40){\circle*{3}}
\put(185,3){$B$}
\end{picture}
\end{center}
\caption{Topologies of two-loop Feynman graphs.}
\end{figure}
The two-loop Feynman diagrams involve three- and four-point
vertices, see Fig.\ 1.
Therefore it is sufficient to expand the classical action up to
the fourth order in quantum superfields,
\bea
S&=&S_2+S_{\rm int}+\ldots
\label{Ssplit}\,,\\
S_2&=&-\int d^3x d^4\theta\, K''_{\Phi\bar\Phi}\phi\bar\phi
-\left(\int d^3x d^2\theta\, \frac12W'' \phi^2 +c.c.\right)\,,
\label{S2}\\
S_{\rm int}&=&-\frac12\int d^3x d^4\theta\left(
K^{(3)}_{\Phi^2\bar\Phi}\phi^2\bar\phi
+K^{(3)}_{\bar\Phi^2\Phi}\bar\phi^2\phi
+\frac12 K^{(4)}_{\Phi^2\bar\Phi^2}\phi^2\bar\phi^2
+\frac13 K^{(4)}_{\Phi^3\bar\Phi}\phi^3\bar\phi
+\frac13 K^{(4)}_{\bar\Phi^3\Phi}\bar\phi^3\phi
\right)\nn\\&&
-\left(\frac16\int d^3x d^2\theta[W'''\phi^3+\frac14 W^{(4)}\phi^4]
 +c.c.\right)\,.
\label{Sint}
\eea
In this decomposition we assume that the background fields are
constant, (\ref{const-back}). Therefore we omitted the
terms involving $\bar D^2 K^{(n)}_{\Phi^n}$ and $D^2 K^{(n)}_{\bar\Phi^n}$ since they do not
contribute to the effective K\"ahler potential.

The quadratic action (\ref{S2}) defines the propagator which can
be explicitly written for the constant field background,
\be
\left(
\begin{array}{cc}
G_{++}(z_1,z_2) & G_{+-}(z_1,z_2)\\
G_{-+}(z_1,z_2) & G_{--}(z_1,z_2)
\end{array}
\right)
=\left(
\begin{array}{cc}
\frac{\bar W''}{(K''_{\Phi\bar\Phi})^2}
\frac1{\square+M^2}\delta_+(z_1,z_2)
&
-\frac{1}{4K''_{\Phi\bar\Phi}}\frac1{\square+M^2}
\bar D^2\delta_-(z_1,z_2) \\
-\frac{1}{4K''_{\Phi\bar\Phi}}\frac1{\square+M^2}
 D^2\delta_+(z_1,z_2) &
-\frac{ W''}{(K''_{\Phi\bar\Phi})^2}
\frac1{\square+M^2}\delta_-(z_1,z_2)
\end{array}
\right).
\label{prop}
\ee
The interaction vertices can be read from (\ref{Sint}).

We follow standard procedure of computing the two-loop Feynman diagrams
 in supersymmetric field theories \cite{book}. Each line at Fig.\
1 corresponds to one of the elements of the matrix propagators
(\ref{prop}).
Each vertex can be either (anti)chiral, e.g.,
\be
S'''_{\Phi^3}(z_1,z_2,z_3)\equiv
\frac{\delta^3 S}{\delta \Phi(z_1)\delta\Phi(z_2)\delta\Phi(z_3)}=
-W'''(z_3)\delta_+(z_1,z_2)\delta_+(z_2,z_3)\,,
\label{S3}
\ee
or, non-chiral, e.g.,
\bea
S'''_{\Phi^2\bar\Phi}(z_1,z_2,z_3)\equiv
\frac{\delta^3 S}{\delta \Phi(z_1)\delta\Phi(z_2)\delta\bar\Phi(z_3)}&=&\frac14
D^2_{(3)}[K'''_{\Phi^2\bar\Phi}(z_3)\delta_+(z_2,z_3)]\delta_+(z_1,z_2)\,,
\label{S3-2}\\
S'''_{\Phi\bar\Phi^2}(z_1,z_2,z_3)\equiv
\frac{\delta^3 S}{\delta \Phi(z_1)\delta\bar\Phi(z_2)\delta\bar\Phi(z_3)}&=&\frac14
D^2_{(2)}[
K'''_{\Phi\bar\Phi^2}(z_2)\delta_+(z_1,z_2)]\delta_-(z_2,z_3)\,.
\label{S3-1}
\eea
Each non-chiral vertex has extra $D^2$ or $\bar D^2$ operator in comparison
with the chiral one. As a consequence, there are many combinatoric
possibilities with these propagators and vertices to construct the
two-loop diagrams shown at Fig.\ 1. However, not all these
diagrams contribute to the effective K\"ahler potential. Only
those diagrams are eligible which have specific number of the $D^2$ and $\bar
D^2$ operators which are necessary for restoring full superspace measures at each
vertex using (\ref{measure}) and then applying the identity
(\ref{id}) for each loop. Analyzing the diagrams of topology $A$
at Fig.\ 1, one can see that only one such diagram contributes,
\be
\Gamma_A=-\frac12\int d^5z_1 d^5\bar z_2 d^5z_3 d^5\bar z_4\,
S^{(4)}_{\Phi^2\bar\Phi^2}(z_1, z_2,z_3, z_4)
 G_{+-}(z_1, z_2)G_{+-}(z_3, z_4)\,.
\label{GA}
\ee
Here $S^{(4)}$ is the fourth variational derivative of the
classical action written explicitly in (\ref{S4}).
Among the diagrams of the topology $B$ contributing to the
effective K\"ahler potential, there are five various
terms,
\bea
\Gamma_{B_1}&=&-\frac12\int d^5z_1 d^5\bar z_2 d^5\bar z_3 d^5z_4d^5z_5 d^5z_6\,
S^{(3)}_{\Phi\bar\Phi^2}(z_1,z_2,z_3)S^{(3)}_{\Phi^3}(z_4,z_5,z_6)
\nn\\&&\times
G_{++}(z_1,z_4)G_{-+}( z_2,z_5)G_{-+}(z_3,z_6)\,,\label{B1}
\\
\Gamma_{B_2}&=&
-\frac16\int d^5\bar z_1 d^5\bar z_2 d^5\bar z_3 d^5z_4d^5z_5 d^5z_6\,
S^{(3)}_{\bar\Phi^3}(z_1,z_2,z_3)S^{(3)}_{\Phi^3}(z_4,z_5,z_6)
\nn\\&&\times
G_{-+}(z_1,z_4)G_{-+}(z_2,z_5)G_{-+}( z_3,z_6)\,,\label{B2}
\\
\Gamma_{B_3}&=&
-\int d^5z_1 d^5\bar z_2 d^5\bar z_3 d^5z_4d^5z_5 d^5\bar z_6\,
S^{(3)}_{\Phi\bar\Phi^2}(z_1,z_2,z_3)
S^{(3)}_{\Phi^2\bar\Phi}(z_4,z_5,z_6)
\nn\\&&\times
G_{++}(z_1,z_4)G_{-+}( z_2,z_5)G_{--}( z_3, z_6)\,,\label{B3}
\\
\Gamma_{B_4}&=&-\frac12\int d^5z_1 d^5\bar z_2 d^5\bar z_3 d^5z_4d^5z_5 d^5\bar z_6\,
S^{(3)}_{\Phi\bar\Phi^2}(z_1,z_2,z_3)
S^{(3)}_{\Phi^2\bar\Phi}(z_4,z_5,z_6)
\nn\\&&\times
G_{+-}(z_1, z_6)G_{-+}( z_2,z_5)G_{-+}( z_3,z_4)\,,\label{B4}
\\
\Gamma_{B_5}&=&
-\frac12\int d^5\bar z_1 d^5\bar z_2 d^5\bar z_3 d^5z_4d^5z_5 d^5\bar z_6\,
S^{(3)}_{\bar\Phi^3}(z_1,z_2,z_3)
S^{(3)}_{\Phi^2\bar\Phi}(z_4,z_5,z_6)
\nn\\&&\times
G_{-+}(z_1,z_4)G_{-+}(z_2,z_5)G_{--}(z_3,z_6)\,.
\label{B5}
\eea

The details of the computations of (\ref{GA})--(\ref{B5}) are
given in the Appendix \ref{A1}. The resulting two-loop contributions to
the K\"ahler potential $K^{(2)}_{\rm eff}$ can be written as the
sum of the divergent and finite parts,
\be
K^{(2)}=K^{(2)}_{\rm div}+K^{(2)}_{\rm fin}\,,
\ee
where
\bea
K^{(2)}_{\rm div}&=&\frac1{64\pi^2\epsilon}
\bigg[3\frac{|K'''_{\Phi\bar\Phi^2}W''|^2}{(K''_{\Phi\bar\Phi})^5}
+\frac13\frac{|W'''|^2}{(K''_{\Phi\bar\Phi})^3}
-\frac{K'''_{\Phi\bar\Phi^2}W'''\bar W''+K'''_{\Phi^2\bar\Phi}W''\bar W'''}{(K''_{\Phi\bar\Phi})^4}
\bigg]\,,\\
K^{(2)}_{\rm fin}&=&\frac1{32\pi^2}\bigg[
\frac{K^{(4)}_{\Phi^2\bar\Phi^2}|W''|^2}{(K''_{\phi\bar\phi})^4}
+\frac12(\gamma+\ln\frac{M^2}{\mu^2})
\frac{K'''_{\Phi\bar\Phi^2}W'''\bar W''+K'''_{\Phi^2\bar\Phi}W''\bar W'''}{
 (K''_{\Phi\bar\Phi})^4}
\nn\\&&
-\frac16(\gamma+\ln\frac{M^2}{\mu^2})\frac{|W'''|^2}{(K''_{\Phi\bar\Phi})^3}
-(1+\frac32\gamma+\frac32\ln\frac{M^2}{\mu^2})
\frac{|K'''_{\Phi\bar\Phi^2}W''|^2}{(K''_{\Phi\bar\Phi})^5}
\bigg]\,,
\label{2loopresult}
\eea
where $M^2=|W''|^2/(K''_{\Phi\bar\Phi})^2$, $\mu^2$ is a
normalization point and $\epsilon$ is the parameter of dimensional
regularization, $d=3-2\epsilon$.

\section{Effective chiral potential}
\setcounter{equation}{0}
\subsection{General properties}

For the four-dimensional $\cN=1$ supersymmetric field theories the
non-renormalization theorem says that any contribution to the
effective action can be represented by an expression in the full
superspace (see e.g. \cite{GGRS,book}), \be \int d^4\theta \,
f(\Phi,\bar\Phi)\,, \label{non-ren} \ee with $f(\Phi,\bar\Phi)$
being some function of superfields (with or without derivatives).
One can easily check that all steps of proof of this theorem and
general conclusion remain to be true for the ${\cal N}=2, d=3$
models as well.

Naively, one can think that (\ref{non-ren}) forbids any quantum
contributions to the chiral effective potential $W_{\rm
eff}$, but it is well-known
for the four-dimensional chiral
superfield model \cite{West1,West2,BKP1,BKP2,BP1,BP2}
that the effective action acquires finite quantum corrections
in the full superspace of the form
\be
\int d^4\theta\,f(\Phi)\left(
-\frac{D^2}{4\square}
\right)g(\Phi)\,,
\label{non-local}
\ee
which do not contradict the non-renormalization theorem. Here $f$
and $g$ are some functions. In the present section we
demonstrate that for the three-dimensional chiral superfield model
the two-loop Feynman diagrams also contain the terms of the form
(\ref{non-local}) which result in the contributions to the chiral
effective potential,
\be
\int d^3x d^4\theta\,f(\Phi)\left(
-\frac{D^2}{4\square}
\right)g(\Phi)
=\int d^3x d^2\theta\, f(\Phi)g(\Phi)\,.
\label{no-nonren}
\ee

It is important to note that the terms in the effective action
like (\ref{non-local}) can appear only in massless models. Indeed,
the propagator in a massive model involves the operator
$(\square+m^2)^{-1}$ instead of $\square^{-1}$. As a consequence,
in the massive model the relation (\ref{no-nonren}) gets modified
as
\be
\int d^3x d^4\theta\,f(\Phi)\left(
-\frac{D^2}{4(\square+m^2)}
\right)g(\Phi)=\int d^3x d^2\theta\,
f(\Phi)\left(
\frac{\square}{\square+m^2}
\right)g(\Phi)\,,
\ee
but this expression vanishes in the limit of slowly
varying fields unless $m=0$. Hence, a non-trivial chiral effective
potential may be present only in the massless theory. This conclusion is
completely analogous to the one for the four-dimensional $\cN=1$
chiral superfield model \cite{BKP1,BKP2,BP1,BP2}. Therefore, in
this section we assume that the chiral potential
$W(\Phi)$ in the classical action (\ref{class}) obeys the constraint
\be
W''|_{\Phi=0}=0\,.
\ee

It is important to note that for deriving the terms like
(\ref{non-local}) we cannot employ the constant field
approximation any more.\footnote{Recall that the propagator (\ref{prop})
was derived in the constant field approximation.} The
background chiral superfield $\Phi$ should be arbitrary throughout
the computations while the anti-chiral one can be freely sent to
zero,
\be
\bar\Phi=0\,,\qquad \Phi \mbox{ is arbitrary.}
\label{back}
\ee
Only after computing all momentum integrals and passing to the
chiral subspace using (\ref{no-nonren}), one can apply the constant
field approximation to single out the contributions to the chiral
effective potential.

\subsection{Analysis of possible Feynman diagrams contributing
to chiral effective potential}

The relation (\ref{no-nonren}) shows that
only those Feynman diagrams contribute to the chiral effective
potential which contain one $D^2$ operator on the external lines
after performing all the $D$-algebra and one $\square^{-1}$ operator
after computing the momentum integrals. First of all, we point out
that the one-loop diagrams cannot contribute to the effective
chiral potential as soon as the corresponding momentum integral
yields only the odd power of external momenta $|p|=\sqrt{p^m p_m}$,
\be
\int d^3k\frac1{k^2}\frac1{(k+p)^{2n}}\propto\frac1{|p|^{2n-1}}\,.
\ee
Because of (\ref{box}), the $D$-algebra produces only even powers of momenta, $p^{2n}$.
Therefore, for the rest of this section we will consider two-loop diagrams
only.

As in the previous section, we split the chiral field $\Phi$ into
the ``background'' $\Phi$ and ``quantum'' $\phi$ parts,
$\Phi\to\Phi+\phi$, and consider the decomposition of the classical
action (\ref{class}) up to the fourth order in the quantum
superfields (c.f. (\ref{Ssplit})--(\ref{Sint})),
\bea
S&=&S_2+S_{\rm int,1}+S_{\rm int,2}+S_{\rm int,3}+\ldots\,,\\
S_2&=&-\int d^3x d^4\theta\, K''_{\Phi\bar\Phi}\phi\bar\phi\,,
\label{S2_}\\
S_{\rm int,1}&=&-\frac12\int d^3x d^4\theta\bigg(
K^{(3)}_{\Phi^2\bar\Phi}\phi^2\bar\phi
+K^{(3)}_{\bar\Phi^2\Phi}\bar\phi^2\phi
+\frac12 K^{(4)}_{\Phi^2\bar\Phi^2}\phi^2\bar\phi^2
\nn\\&&
+\frac13 K^{(4)}_{\Phi^3\bar\Phi}\phi^3\bar\phi
+\frac13 K^{(4)}_{\Phi\bar\Phi^3}\bar\phi^3\phi
\bigg)\,,\label{Sint1}\\
S_{\rm int,2}&=&-\frac12\int d^3x d^2\theta\left(
\phi^2(-\frac14\bar D^2)K''_{\Phi^2}
+\frac13\phi^3(-\frac14\bar D^2)K'''_{\Phi^3}
+\frac1{12}\phi^4(-\frac14\bar D^2)K^{(4)}_{\Phi^4}
\right)\nn\\
&&-\frac12\int d^3x d^2\bar\theta\left(
\bar\phi^2(-\frac14 D^2)K''_{\bar\Phi^2}
+\frac13\bar\phi^3(-\frac14 D^2)K'''_{\bar\Phi^3}
+\frac1{12}\bar\phi^4(-\frac14 D^2)K^{(4)}_{\bar\Phi^4}
\right)\,,\label{Sint2}\\
S_{\rm int,3}&=&-\frac12\int d^3x d^2\theta\left(
\phi^2 W''+\frac13\phi^3 W'''+\frac1{12}\phi^4W^{(4)}
\right)+c.c.\label{Sint3}
\eea
We point out that, owing to (\ref{back}), all derivatives of $K$ and $W$, $\bar W$ in this
expansion are considered at vanishing antichiral background field $\bar\Phi$.
Therefore all these factors are chiral.

The action $S_2$ is responsible for the propagator,
\be
\langle\phi(z)\bar\phi(z')\rangle\equiv G_0(z,z')=-\frac1{K''_{\Phi\bar\Phi}}\frac{D^2}{4\square}
 \delta_+(z,z')
=\frac1{K''_{\Phi\bar\Phi}}\frac{D^2 \bar D^2}{16\square}
 \delta^7(z-z')
 \,.
\label{prop1}
\ee
Note that this propagator contains four Grassmann derivatives
on the delta-function while the propagators
$\langle\phi\phi\rangle$ and $\langle\bar\phi\bar\phi\rangle$
are traded for the corresponding vertices.\footnote{The terms in the actions (\ref{Sint2}),
(\ref{Sint3}) containing
$\phi^2$ and $\bar\phi^2$ are treated as vertices rather than
the propagators for the quantum fields.}
 This restricts the number of possible Feynman
diagrams with these propagators.

\begin{table}[tb]
\begin{center}
\begin{tabular}{l|l||l|l||l|l}
\multicolumn{2}{c||}{Full superspace} &
\multicolumn{2}{c||}{Antichiral vertices}&
\multicolumn{2}{c}{Chiral vertices}\\\hline\hline
$K^{(3)}_{\Phi^2\bar\Phi}\phi^2\bar\phi$ &
\begin{picture}(40,40)
\thicklines
\put(15,15){\line(1,0){15}}
\put(15,15){\line(-1,0){15}}
\put(15,15){\line(0,-1){15}}
\put(15,15){\line(0,1){15}}
\put(15.5,15){\line(0,1){15}}
\put(14.5,15){\line(0,1){15}}
\put(15,15){\circle*{3}}
\put(17,22){\scriptsize$K^{(3)}_{\Phi^2\bar\Phi}$}
\end{picture}&
$\bar W'''\bar\phi^3$ &
\begin{picture}(40,40)
\thicklines
\put(15,15){\line(1,0){15}}
\put(15,15){\line(-1,0){15}}
\put(15,15){\line(0,-1){15}}
\put(15,15){\circle*{5}}
\put(10,20){\scriptsize$\bar W'''$}
\end{picture}&
$W''\phi^2$ &
\begin{picture}(40,40)
\thicklines
\put(15,10){\line(1,0){15}}
\put(15,10){\line(-1,0){15}}
\put(15,10){\line(0,1){15}}
\put(15.5,10){\line(0,1){15}}
\put(14.5,10){\line(0,1){15}}
\put(15,10){\circle{5}}
\put(18,17){\scriptsize$W''$}
\end{picture} \\\hline
$K^{(3)}_{\bar\Phi^2\Phi}\bar\phi^2\phi$ &
\begin{picture}(40,40)
\thicklines
\put(15,15){\line(1,0){15}}
\put(15,15){\line(-1,0){15}}
\put(15,15){\line(0,-1){15}}
\put(15,15){\line(0,1){15}}
\put(15.5,15){\line(0,1){15}}
\put(14.5,15){\line(0,1){15}}
\put(15,15){\circle*{3}}
\put(17,22){\scriptsize$K^{(3)}_{\bar\Phi^2\Phi}$}
\end{picture}&
$\bar W^{(4)}\bar\phi^4$ &
\begin{picture}(40,40)
\thicklines
\put(15,15){\line(1,0){15}}
\put(15,15){\line(-1,0){15}}
\put(15,15){\line(1,-1){10}}
\put(15,15){\line(-1,-1){10}}
\put(15,15){\circle*{5}}
\put(10,20){\scriptsize$\bar W^{(4)}$}
\end{picture}&
$W'''\phi^3$ &
\begin{picture}(40,40)
\thicklines
\put(15,15){\line(1,0){15}}
\put(15,15){\line(-1,0){15}}
\put(15,15){\line(0,1){15}}
\put(15,15){\line(0,-1){15}}
\put(15.5,15){\line(0,1){15}}
\put(14.5,15){\line(0,1){15}}
\put(15,15){\circle{5}}
\put(18,22){\scriptsize$W'''$}
\end{picture} \\\hline
$K^{(4)}_{\bar\Phi^2\Phi^2}\bar\phi^2\phi^2$ &
\begin{picture}(40,40)
\thicklines
\put(15,15){\line(1,0){15}}
\put(15,15){\line(-1,0){15}}
\put(15,15){\line(-1,-1){10}}
\put(15,15){\line(1,-1){10}}
\put(15,15){\line(0,1){15}}
\put(15.5,15){\line(0,1){15}}
\put(14.5,15){\line(0,1){15}}
\put(15,15){\circle*{3}}
\put(17,22){\scriptsize$K^{(4)}_{\bar\Phi^2\Phi^2}$}
\end{picture}&
$(-\frac14 D^2 K''_{\bar\Phi^2})\bar\phi^2$ &
\begin{picture}(40,40)
\thicklines
\put(15,10){\line(1,0){15}}
\put(15,10){\line(-1,0){15}}
\put(15,10){\line(0,1){15}}
\put(15.5,10){\line(0,1){15}}
\put(14.5,10){\line(0,1){15}}
\put(15,10){\circle*{5}}
\put(18,17){\scriptsize$D^2 K''_{\bar\Phi^2}$}
\end{picture}&
$W^{(4)}\phi^4$ &
\begin{picture}(40,40)
\thicklines
\put(15,15){\line(1,0){15}}
\put(15,15){\line(-1,0){15}}
\put(15,15){\line(0,1){15}}
\put(15,15){\line(-1,-1){10}}
\put(15,15){\line(1,-1){10}}
\put(15.5,15){\line(0,1){15}}
\put(14.5,15){\line(0,1){15}}
\put(15,15){\circle{5}}
\put(18,22){\scriptsize$W^{(4)}$}
\end{picture} \\\hline
$K^{(4)}_{\Phi^3\bar\Phi}\phi^3\bar\phi$ &
\begin{picture}(40,40)
\thicklines
\put(15,15){\line(1,0){15}}
\put(15,15){\line(-1,0){15}}
\put(15,15){\line(-1,-1){10}}
\put(15,15){\line(1,-1){10}}
\put(15,15){\line(0,1){15}}
\put(15.5,15){\line(0,1){15}}
\put(14.5,15){\line(0,1){15}}
\put(15,15){\circle*{3}}
\put(17,22){\scriptsize$K^{(4)}_{\Phi^3\bar\Phi}$}
\end{picture}&
$(-\frac14 D^2 K'''_{\bar\Phi^3})\bar\phi^3$ &
\begin{picture}(40,40)
\thicklines
\put(15,15){\line(1,0){15}}
\put(15,15){\line(-1,0){15}}
\put(15,15){\line(0,1){15}}
\put(15,15){\line(0,-1){15}}
\put(15.5,15){\line(0,1){15}}
\put(14.5,15){\line(0,1){15}}
\put(15,15){\circle*{5}}
\put(18,22){\scriptsize$D^2 K'''_{\bar\Phi^3}$}
\end{picture} \\\hline
$K^{(4)}_{\Phi\bar\Phi^3}\phi\bar\phi^3$ &
\begin{picture}(40,40)
\thicklines
\put(15,15){\line(1,0){15}}
\put(15,15){\line(-1,0){15}}
\put(15,15){\line(-1,-1){10}}
\put(15,15){\line(1,-1){10}}
\put(15,15){\line(0,1){15}}
\put(15.5,15){\line(0,1){15}}
\put(14.5,15){\line(0,1){15}}
\put(15,15){\circle*{3}}
\put(17,22){\scriptsize$K^{(4)}_{\Phi\bar\Phi^3}$}
\end{picture}&
$(-\frac14 D^2 K^{(4)}_{\bar\Phi^4})\bar\phi^4$ &
\begin{picture}(40,40)
\thicklines
\put(15,15){\line(1,0){15}}
\put(15,15){\line(-1,0){15}}
\put(15,15){\line(0,1){15}}
\put(15,15){\line(-1,-1){10}}
\put(15,15){\line(1,-1){10}}
\put(15.5,15){\line(0,1){15}}
\put(14.5,15){\line(0,1){15}}
\put(15,15){\circle*{5}}
\put(17,22){\scriptsize$D^2 K^{(4)}_{\bar\Phi^4}$}
\end{picture} \\\hline
\end{tabular}
\end{center}
\caption{Graphical representations of the vertices which are relevant
for two-loop computations. Thick lines stand for the expressions
depending on the background superfields while the thin ones mean
the quantum superfields.}
\label{table1}
\end{table}

The actions (\ref{Sint1}), (\ref{Sint2}) and (\ref{Sint3}) are
responsible for the vertices which are relevant for two-loop Feynman
graphs. We use the graphical representations of these vertices
according to Table \ref{table1}. It is convenient to distinguish these
vertices with respect to the type of superspace over which they are
integrated (chiral, antichiral or full $\cN=2$, $d=3$ superspace).
Let us comment on each of these types in more details.

The action $S_{\rm int,1}$ yields the vertices
in the full superspace which involve $K^{(3)}_{\Phi^2\bar\Phi}\phi^2\bar\phi$,
$K^{(3)}_{\bar\Phi^2\Phi}\bar\phi^2\phi$,
$K^{(4)}_{\Phi^2\bar\Phi^2}\phi^2\bar\phi^2$,
$\frac13 K^{(4)}_{\Phi^3\bar\Phi}\phi^3\bar\phi$ and
$\frac13 K^{(4)}_{\bar\Phi^3\Phi}\bar\phi^3\phi$. Such vertices
bring extra $D^2$ or $\bar D^2$ operators as compared with the
(anti)chiral vertices, see, e.g.,
(\ref{S3-2}) and (\ref{S3-1}). These $D^2$-operators can
hit both external and internal lines in a diagram. Once they hit
the external lines, they do not affect the loop momenta any more.
In fact, we need to consider only those diagrams which
have only one $D^2$ operator on the external lines that is
necessary for chiral contributions due to (\ref{no-nonren}). If these
operators hit the internal lines, they either can be used to
restore full superspace measure (\ref{measure}) or increase the
superficial degree of divergence of the diagram owing to the
$D$-algebra (\ref{box}). In the following, by $n_1$ we denote the
number of vertices in a Feynman diagram corresponding to the
action $S_{\rm int,1}$.

The action $S_{\rm int,2}$ contains chiral and antichiral
vertices. Note that the chiral vertices bear the $\bar D^2$
operators on the external lines, but we need $D^2$ operator to
apply the identity (\ref{no-nonren}). Hence, we can neglect all
the chiral vertices given in the first line of (\ref{Sint2}) and
consider only the antichiral ones in the second line.
By $n_2$ we denote the number of vertices corresponding to the
second line of (\ref{Sint2}) in a Feynman diagram.
It is
clear that a diagram contributing to the chiral effective
potential should contain no more than one such vertex,
$n_2=0$ or $n_2=1$.

The action $S_{\rm int,3}$ is responsible for the (anti)chiral
vertices without $D^2$ or $\bar D^2$ operators. Denote the number
of such vertices by $n_3$.

Our aim now is to analyze the two-loop diagrams at Fig.~1 and to
fix the numbers $n_1$, $n_2$ and $n_3$ which correspond to
non-trivial contributions to the chiral effective potential.
The strategy of our considerations is as follows:
\begin{itemize}
\item Draw all admissible two-loop diagrams with the propagator
(\ref{prop1}) and with the vertices given in Table \ref{table1};
\item Restore full superspace measures at all (anti)chiral
vertices using the Grassmann derivatives from the propagators;
\item All other Grassmann derivatives can be integrated by parts
producing a number of different terms, but we need to consider
only those of them which contain exactly two derivatives $D_\alpha$ on
the external lines;
\item Two operators $\bar D^2 D^2$ are eaten by the identity
(\ref{id})(by one for each loop);
\item The remaining Grassmann derivatives generate the internal
momenta which increases the superficial degree of divergence of
diagrams;
\item Only those diagrams are eligible in which the momentum loop
integrals produce the power of external momenta as $p^{-2}$. Then,
upon application of the identity (\ref{no-nonren}), this diagram
contributes to the effective chiral potential.
\end{itemize}

\subsubsection{Diagrams of topology $A$}
The diagrams of topology $A$ at Fig.~1 have one of the following
quartic vertices, $K^{(4)}_{\Phi^2\bar\Phi^2}\phi^2\bar\phi^2$,
$\frac13 K^{(4)}_{\Phi^3\bar\Phi}\phi^3\bar\phi$,
$\frac13 K^{(4)}_{\bar\Phi^3\Phi}\bar\phi^3\phi$,
$\bar\phi^4(-\frac14 D^2)K^{(4)}_{\bar\Phi^4}$,
$\phi^4W^{(4)}$ or $\bar\phi^4\bar W^{(4)}$.
Consider a Feynman graph of this topology which involves
$n_2$ vertices with
$\bar\phi^2(-\frac14 D^2)K''_{\bar\Phi^2}$ and $n_3$ vertices with
$\phi^2 W''$. Recall that there is no antichiral $\bar\phi^2 \bar W''$ vertex
since it vanishes for the considered background (\ref{back})
in the massless theory.
Clearly, this diagram should have $n_2+n_3+2$ propagators
(\ref{prop1}) each of which brings the operator $D^2\bar D^2\square^{-1}$.

The non-chiral vertices $K^{(4)}_{\Phi^2\bar\Phi^2}\phi^2\bar\phi^2$,
$\frac13 K^{(4)}_{\Phi^3\bar\Phi}\phi^3\bar\phi$,
$\frac13 K^{(4)}_{\bar\Phi^3\Phi}\bar\phi^3\phi$ are integrated over
the full superspace. Therefore, we need to restore the full
superspace measure only for the $n_2$ antichiral vertices with
$\bar\phi^2(-\frac14 D^2)K''_{\bar\Phi^2}$ and for $n_3$ chiral vertices with
$\phi^2 W''$ with the help of the $D$-operators which are present in the
propagators. Moreover, doing integration by parts we have to collect
$1-n_2$ operators $D^2$ on the
external lines which are required for the identity (\ref{no-nonren}). For each loop
we have to apply the identity (\ref{id}) which eats the $D^2\bar
D^2$ operator. As a result, we are left with $(n_3+n_2-1)$
operators $D^2$ and with $n_2$  operators $\bar D^2$. We cannot
put any more Grassmann derivatives on the external
lines as we wish to get the chiral effective potential.
Hence, all these Grassmann derivatives should recombine into the
internal momenta which means that there should be equal numbers of $D^2$
and $\bar D^2$ operators, $n_3+n_2-1=n_2$. As a result, the
eligible diagrams have only $n_3=1$ vertices $\phi^2 W''$. But
for the number $n_2$ the only possibility is $n_2=1$ sine for $n_2=0$ the
two-loop diagram with the only external line vanishes
automatically. After using all the $D$-operators from the
propagators as is described here, we are left with one $D^2$ and
one $\bar D^2$ which produce one $\square$ operator in the nominator,
but four box operators
stand in the denominators of four propagators. The resulting momentum integral
is something like
\be
\int \frac{d^3k_1 d^3 k_2\, k_1^2}{k_1^2
k_2^2(k_1+p)^2(k_2+p)^2}\propto p^0\,,
\label{int}
\ee
but we need $p^{-2}\to \square^{-1}$  to apply (\ref{no-nonren}).
Hence, the diagrams with $K^{(4)}_{\Phi^2\bar\Phi^2}\phi^2\bar\phi^2$,
$\frac13 K^{(4)}_{\Phi^3\bar\Phi}\phi^3\bar\phi$,
$\frac13 K^{(4)}_{\bar\Phi^3\Phi}\bar\phi^3\phi$ vertices do not
contribute to the effective chiral potential.

Consider a diagram with $\bar\phi^4(-\frac14
D^2)K^{(4)}_{\bar\Phi^4}$ vertex which is integrated over the
antichiral superspace. Here we already have one $D^2$ on the
external line, hence no vertices with $\bar\phi^2(-\frac14
D^2)K''_{\bar\Phi^2}$ are allowed, $n_2=0$. Since we have only
$\langle\phi\bar\phi\rangle$ propagators but neither
$\langle\phi\phi\rangle$ nor $\langle\bar\phi\bar\phi\rangle$, the
only possibility to build the diagram of the topology $A$ is by
attaching one $\phi^2 W''$ vertex for each loop, i.e., $n_3=2$.
>From four propagators we take one $D^2$ operator and two $\bar
D^2$ operators to restore the full superspace measure at each vertex and we
are left with unbalanced number of such operators leading to the
null contribution for the effective chiral potential.

Consider now a diagram with the quartic vertex $\phi^4W^{(4)}$.
It is easy to see that such a diagram should have at least two
$\bar\phi^2(-\frac14 D^2)K''_{\bar\Phi^2}$ vertices because we
have only $\langle\phi\bar\phi\rangle$ propagator. But this leads
to two $D^2$ operators on the external lines while we need only
one to apply (\ref{no-nonren}). Hence, such diagrams do not
contribute to the effective chiral potential.

Finally, we have to consider a diagram with the $\bar\phi^4\bar
W^{(4)}$ quartic vertex. Assume that it involves also $n_2$ vertices with
$\bar\phi^2(-\frac14 D^2)K''_{\bar\Phi^2}$ and $n_3$ vertices with
$\phi^2 W''$. Clearly, there are $n_2+n_3+2$ propagators each of
which brings the $\square^{-1}D^2\bar D^2$ operator. We need
to put on the external lines $1-n_2$ operators $D^2$ to satisfy
(\ref{no-nonren}). To restore full superspace measure,
we need also $n_2+1$ operators $D^2$ and $n_3$ operators $\bar
D^2$. One $D^2\bar D^2$ operator is eaten by each loop because of
the identity (\ref{id}). As a result we are left with
$n_2+n_3-2$ operators $D^2$ and $n_2$ operators $\bar D^2$. These
numbers should be equal since we cannot put these derivatives on
the external lines, $n_2+n_3-2=n_2$. Hence, the only possibility
is the diagram with $n_2=0$ and $n_3=2$. The corresponding
momentum integral has exactly right power of external momenta to
apply (\ref{no-nonren}),
\be
\int \frac{d^3k_1 d^3k_2}{k_1^2
k_2^2(p+k_1)^2(p+k_2)^2}=\frac{\pi^6}{p^2}\,.
\ee
As a result, the diagram given at Fig.~2a can contribute to the
effective chiral potential.

\subsubsection{Diagrams of topology $B$}
There are two non-chiral vertices
$K^{(3)}_{\Phi^2\bar\Phi}\phi^2\bar\phi$ and
$K^{(3)}_{\bar\Phi^2\Phi}\bar\phi^2\phi$ which are integrated in
the full superspace, two chiral vertices $\phi^3 W'''$,
$\phi^3(-\frac14\bar D^2)K'''_{\Phi^3}$ and two antichiral ones,
$\bar \phi^3 \bar W'''$, $\bar \phi^3(-\frac14 D^2)K'''_{\bar
\Phi^3}$. Hence, there are many different possibilities to
construct the two-loop diagrams of topology $B$ with these
vertices. Let us analyze all of them.

Consider a diagram with two vertices in the full superspace,
either $K^{(3)}_{\Phi^2\bar\Phi}\phi^2\bar\phi$ or
$K^{(3)}_{\bar\Phi^2\Phi}\bar\phi^2\phi$ and with $n_2$ vertices
$\bar\phi^2 (-\frac14 D^2)K''_{\bar\Phi^2}$ and with $n_3$
vertices $\phi^2 W''$. From the propagators we have $3+n_2+n_3$ operators
$\square^{-1}D^2\bar D^2$ out of which we use $n_2$
operators $D^2$ and $n_3$ operators $\bar D^2$ to restore full
superspace measures. Two operators $D^2\bar D^2$ are eaten owing
to the identity (\ref{id}), one for each loop. And we need to put
$1-n_2$ operators $D^2$ on the external lines to apply
(\ref{no-nonren}). As a result, we are left with
$n_2+n_3$ operators $D^2$ and with $n_2+1$ operators $\bar D^2$.
These numbers should coincide, $n_2+n_3=n_2+1$, to produce the
internal momenta in the corresponding power. Hence, $n_3=1$ and
$n_2=0$ or $n_2=1$. For both these values of $n_2$, the momentum
integral results in the same power of external momenta as in
(\ref{int}). But we need one $\square^{-1}$ after the integration
over the loop momenta to apply (\ref{no-nonren}). Hence, the diagrams
with only the full superspace cubic vertices do not contribute to
the effective choral potential.

Consider a diagram with one full superspace cubic vertex $K^{(3)}_{\Phi^2\bar\Phi}\phi^2\bar\phi$ or
$K^{(3)}_{\bar\Phi^2\Phi}\bar\phi^2\phi$ and one chiral cubic vertex
$\phi^3 W'''$. As contrasted with the previous case,
we use one $\bar D^2$ operator to restore full superspace measure
in the chiral vertex and we are left $n_2+n_3$ operators $D^2$ and
$n_2$ operators $\bar D^2$. Hence, $n_3=0$ and $n_2=0$ or $n_2=1$.
For both values of $n_2$ the momentum integral has insufficient
power of momenta in the denominator to produce $\square^{-1}$
operator which is necessary for (\ref{no-nonren}). The cubic chiral vertex
$\phi^3 (-\frac14 \bar D^2)K'''_{\Phi^3}$ can be considered in
similar lines with the same negative conclusion.

Consider a diagram with one full superspace cubic vertex
$K^{(3)}_{\Phi^2\bar\Phi}\phi^2\bar\phi$ or
$K^{(3)}_{\bar\Phi^2\Phi}\bar\phi^2\phi$ and one antichiral cubic vertex
$\bar\phi^3 \bar W'''$. In contrast with the previously considered
case, to restore full superspace measure we need the
$\bar D^2$ operator instead of $D^2$. Hence, after contracting
Grassmann loops to points, we are left with $n_2+n_3-1$ operators
$D^2$ and $n_2+1$ operators $\bar D^2$. From the equation
$n_2+n_3-1=n_2+1$ we get $n_3=2$ and $n_2=0$ or $n_2=1$. Take, for
instance, $n_2=0$, then we have five operators $\square$ in
the denominator and one $\square$ in the nominator owing to the
$D$-algebra. The resulting momentum integral gives exactly
$\square^{-1}$ operator on the external lines which is necessary
for the identity (\ref{no-nonren}). The momentum integral gives
the same power for $n_2=1$. Hence, both these diagrams $b_1$ and
$b_3$ at Fig.\ 2 can
contribute to the effective chiral potential. Their calculation
will be performed in the next subsection. Note that similar diagram with
the antichiral cubic vertex $\bar\phi^3(-\frac14
D^2)K'''_{\bar\Phi^3}$ does not contribute.

Let us analyze the diagrams with purely chiral or antichiral
cubic vertices. Take a diagram with two cubic vertices
$\phi^3 W'''$. Now we need to use two
extra $\bar D^2$ operators to restore full superspace measure in
these vertices. As a result we are left with $n_2+n_3$ operators
$D^2$ and $n_2-1$ operators $\bar D^2$. Comparing these numbers,
we get $n_3=-1$ that is impossible. Therefore such diagrams do not
contribute to the effective chiral potential.

Take a diagram with one chiral cubic vertex $\phi^3W'''$
and one antichiral one $\bar\phi^3\bar W'''$. After restoring full superspace measures
at these vertices we are left with $n_3+n_2-1$ operators $D^2$
and $n_2$ operators $\bar D^2$. Comparing these numbers we get
$n_3=1$ and $n_2=1$. The
momentum integral yields the external momenta as $p^{-2}$ and,
hence, this diagram
may contribute to the effective chiral potential. Such a diagram
is given at Fig.\ 2 $b_2$. However,
similar diagram with the $\bar\phi^3(-\frac14
D^2)K'''_{\bar\Phi^3}$ cubic vertex does not contribute.

Consider a diagram with two antichiral vertices
$\bar \phi^3 \bar W'''$ which require two extra $D^2$ operators
for restoring full superspace measure. We are left with
$n_2+n_3-2$ operators $D^2$ and with $n_2+1$ operators $\bar D^2$.
Comparing these numbers we see that $n_3=3$ and $n_2=0$ is
required since we have only $\langle \phi\bar\phi\rangle$
propagator. However, from six propagators we get six $\square^{-1}$
operators, but only one $\square$ operator comes from the remaining
$D$-algebra. The resulting loop momentum integral can produce
$\square^{-2}$ instead of $\square^{-1}$ which is required for
(\ref{no-nonren}). Hence, these diagrams do not contribute to the
effective chiral potential. Note that similar diagram with one
$\bar\phi^3(-\frac14 D^2)K'''_{\bar\Phi^3}$ vertex instead of
$\bar \phi^3 \bar W'''$ also gives vanishing contribution to the
effective chiral potential.

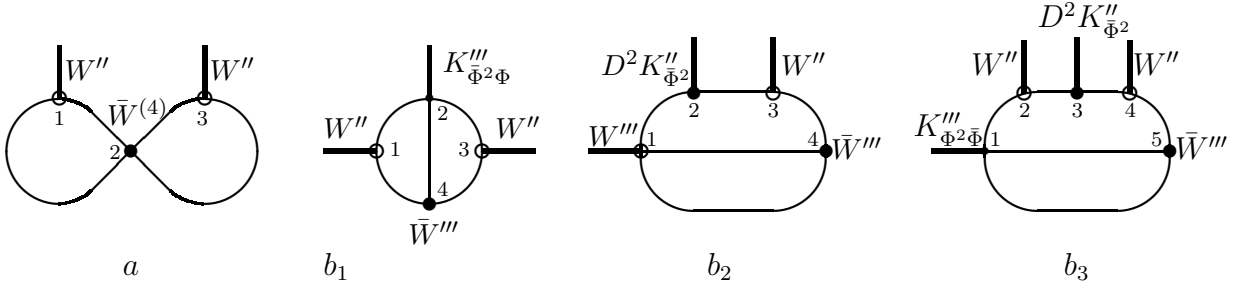
\begin{figure}[tb]
\begin{center}
\begin{picture}(480,100)
\thicklines
\put(30,50){\oval(40,40)[l]}
\put(85,50){\oval(40,40)[r]}
\qbezier(30,30)(38,31)(42,35)
\qbezier(30,70)(38,69)(42,65)
\put(42,35){\line(1,1){30}}
\put(42,65){\line(1,-1){30}}
\qbezier(84,30)(76,31)(72,35)
\qbezier(84,70)(76,69)(72,65)
\put(57,50){\circle*{5}}
\put(30,70){\circle{5}}
\put(85,70){\circle{5}}
\put(30,70){\line(0,1){20}}
\put(85,70){\line(0,1){20}}
\put(29.5,70){\line(0,1){20}}
\put(85.5,70){\line(0,1){20}}
\put(30.5,70){\line(0,1){20}}
\put(84.5,70){\line(0,1){20}}
\put(48,60){\small$\bar W^{(4)}$}
\put(32,77){\small$W''$}
\put(87,77){\small$W''$}
\put(28,60){\scriptsize 1}
\put(49,47){\scriptsize 2}
\put(82,60){\scriptsize 3}
\put(54,3){$a$}
\put(170,50){\circle{40}}
\put(130,50){\line(1,0){20}}
\put(130,50.5){\line(1,0){20}}
\put(130,49.5){\line(1,0){20}}
\put(210,50){\line(-1,0){20}}
\put(210,49.5){\line(-1,0){20}}
\put(210,50.5){\line(-1,0){20}}
\put(170,30){\line(0,1){40}}
\put(170,70){\line(0,1){20}}
\put(169.5,70){\line(0,1){20}}
\put(170.5,70){\line(0,1){20}}
\put(150,50){\circle{5}}
\put(190,50){\circle{5}}
\put(170,70){\circle*{3}}
\put(170,30){\circle*{5}}
\put(175,80){\small$K'''_{\bar\Phi^2\Phi}$}
\put(195,55){\small$W''$}
\put(130,55){\small$W''$}
\put(162,15){\small$\bar W'''$}
\put(173,62){\scriptsize 2}
\put(181,48){\scriptsize 3}
\put(173,33){\scriptsize 4}
\put(155,48){\scriptsize 1}
\put(130,3){$b_1$}
\put(285,50){\oval(70,45)}
\put(250,50){\line(1,0){70}}
\put(250,50){\circle{5}}
\put(320,50){\circle*{5}}
\put(270,72){\circle*{5}}
\put(300,72){\circle{5}}
\put(250,50){\line(-1,0){20}}
\put(270,73){\line(0,1){20}}
\put(300,73){\line(0,1){20}}
\put(250,49.5){\line(-1,0){20}}
\put(269.5,73){\line(0,1){20}}
\put(299.5,73){\line(0,1){20}}
\put(250,50.5){\line(-1,0){20}}
\put(270.5,73){\line(0,1){20}}
\put(300.5,73){\line(0,1){20}}
\put(230,52){\small$W'''$}
\put(322,47){\small$\bar W'''$}
\put(235,79){\small$D^2K''_{\bar\Phi^2}$}
\put(303,79){\small$W''$}
\put(252,52){\scriptsize 1}
\put(268,63){\scriptsize 2}
\put(298,63){\scriptsize 3}
\put(313,52){\scriptsize 4}
\put(275,3){$b_2$}
\put(415,50){\oval(70,45)}
\put(380,50){\line(1,0){70}}
\put(380,50){\circle*{3}}
\put(450,50){\circle*{5}}
\put(395,72){\circle{5}}
\put(435,72){\circle{5}}
\put(415,72){\circle*{5}}
\put(380,50){\line(-1,0){20}}
\put(415,73){\line(0,1){20}}
\put(435,72){\line(0,1){20}}
\put(395,72){\line(0,1){20}}
\put(380,49.5){\line(-1,0){20}}
\put(414.5,73){\line(0,1){20}}
\put(434.5,72){\line(0,1){20}}
\put(394.5,72){\line(0,1){20}}
\put(380,50.5){\line(-1,0){20}}
\put(415.5,73){\line(0,1){20}}
\put(435.5,72){\line(0,1){20}}
\put(395.5,72){\line(0,1){20}}
\put(353,56){$K'''_{\Phi^2\bar\Phi}$}
\put(452,47){$\bar W'''$}
\put(400,97){$D^2K''_{\bar\Phi^2}$}
\put(436,79){$W''$}
\put(375,79){$W''$}
\put(382,52){\scriptsize 1}
\put(393,63){\scriptsize 2}
\put(413,63){\scriptsize 3}
\put(433,63){\scriptsize 4}
\put(443,52){\scriptsize 5}
\put(410,3){$b_3$}
\end{picture}
\end{center}
\caption{Diagrams contributing to the effective chiral potential.}
\end{figure}
To summarize, we need to compute only the diagrams depicted at
Fig.~2 to find the contributions to the effective chiral
potential.

\subsection{Results for two-loop chiral effective potential}
Let us denote the contributions to the effective chiral potential
from the supergraphs in Fig.~2 as
\be
W^{(2)}_{\rm eff}=W_a+W_{b_1}+W_{b_2}+W_{b_3}\,.
\label{WW}
\ee
Here the subscripts in the rhs label the contributions form the
corresponding diagrams.
The computation of these Feynman graphs is a standard
routine in supersymmetric quantum field theory \cite{book} with
the only feature:
In general, such diagrams give non-local contributions to the
effective action, but we need to extract from them only the local
pieces in the chiral sector which are relevant for the effective
chiral potential. This procedure of extracting local parts from
Feynman diagrams is usually referred to as the local limit
\cite{BKP1,BKP2,BP1,BP2}. Essentially, this is the limit when all
the external momenta of a Feynman graph are sent to zero. We point
out that only those Feynman graphs contribute to the effective
chiral potential which are finite in the local limit and for which this limit is unique.
All the
diagrams which are singular in the local limit should be
systematically neglected as they are essentially non-local and do
not contain local parts. As is demonstrated in the Appendix A.2,
only the diagram $a$ at Fig.~2 possesses unique and well defined
local limit and contributes to the effective chiral potential as
\be
\label{WA}
W_a=-\frac1{512}
\frac{ \bar W^{(4)} (W'')^2}{(K''_{\Phi\bar\Phi})^4}\,.
\ee
On the contrary, the diagrams $b_1$, $b_2$ and $b_3$ do not
have uniquely defined local parts and, hence,
in accordance with the definition of chiral effective
potential we have
\be
W_{b_1}=W_{b_2}=W_{b_3}=0\,.
\label{vanish}
\ee
All contributions form these diagrams to the effective action are
essentially non-local and are out of our considerations.

Note also that for the Wess-Zumino model (\ref{WZ}) all
contributions to the effective chiral potential from the diagrams
$b_1$, $b_2$ and $b_3$ vanish identically because
$K''_{\bar\Phi^2}=K'''_{\bar\Phi^2\Phi}=0$ in this case.

\section{Effective potentials in the Wess-Zumino model}
\setcounter{equation}{0}
The computations of the effective K\"ahler and chiral potentials
are done for arbitrary classical K\"ahler and chiral potentials in
the model (\ref{class}). Now we consider an application on these
results for the three-dimensional Wess-Zumino model with $K$ and
$W$ given in (\ref{WZ}).
For simplicity, we restrict ourself to the case with
\be
m=0\,,\qquad\lambda=0\,,
\ee
which corresponds to the scale invariant classical action.

Consider the one- (\ref{K}) and two-loop (\ref{2loopresult})
contributions to the effective K\"ahler potential in the
Wess-Zumino model,
\be
K_{\rm eff}=\Phi\bar\Phi\left(
1+\frac{g}{8\pi}-\frac{g^2\gamma}{192\pi^2}
-\frac{g^2}{96\pi^2}\ln\frac{g\Phi\bar\Phi}{2\mu}
\right)\,.
\label{K-WZ}
\ee
The normalization point $\mu$ can be fixed from the condition
\be
\left.\frac{\partial^2 K_{\rm eff}}{\partial\Phi\partial\bar\Phi}
\right|_{\Phi=\Phi_0,\bar\Phi=\bar\Phi_0}=1\,,
\ee
which implies
\be
-\frac{g}{8\pi}+\frac{g^2(\gamma+4)}{192\pi^2}
+\frac{g^2}{192\pi^2}\ln\frac{
g\Phi_0\bar\Phi_0}{2\mu}=0\,.
\ee
Expressing $\mu$ from this equation and substituting back into
(\ref{K-WZ}) we get
\be
K_{\rm eff}=\Phi\bar\Phi+\frac{g^2}{96\pi^2}
\Phi\bar\Phi\left(2-\ln\frac{\Phi\bar\Phi}{\Phi_0\bar\Phi_0}\right)\,.
\label{K-WZ-eff}
\ee

Since for the Wess-Zumino model the following derivatives of the
classical K\"ahler potential vanish, $K''_{\bar\Phi^2}=0$,
$K'''_{\bar\Phi^2\Phi}=0$, only (\ref{WA}) contributes to the
effective chiral potential, \be W_{\rm eff}=\frac g{24}\Phi^4-\frac{
g^3}{2048}\Phi^4\,. \ee
Note that it is finite similarly as in the
four-dimensional case \cite{West1,West2,BKP1,BKP2}. This quantum
contribution means simply a finite renormalization of the coupling
constant,
\be g\to g'=g(1-\frac3{256}g^2)\,. \label{4.7} \ee
Once
this shift is performed, all two-loop quantum contributions to the
chiral potential are accounted by this coupling $g'$. In the rest
of this section we will use $g$ instead of $g'$, for brevity, assuming that
it already takes the quantum corrections into account.

Let us study the effective scalar potential induced by the quantum
corrections (\ref{K-WZ-eff}). Consider only the constant scalar $\varphi$ and
auxiliary $F$ fields,
\be
\Phi=\varphi+\theta^2 F\,,\qquad
\bar\Phi=\bar\varphi-\bar\theta^2 \bar F\,.
\ee
Eliminating the
auxiliary fields in the classical Wess-Zumino model gives
scale-invariant scalar potential,
\be V_{\rm
class}=\frac{g^2}{36}(\varphi\bar\varphi)^3\,.
\ee
Analogously,
elimination of the auxiliary fields in the effective action with the
effective K\"ahler potential (\ref{K-WZ-eff}) yields the effective
scalar potential,
\be
V_{\rm eff}=\frac{g^2}{36}
(\varphi\bar\varphi)^3\left(1+\frac{g^2}{96\pi^2}\ln\frac{\varphi\bar\varphi}{
\varphi_0\bar\varphi_0} +\mbox{higher loop corrections}\right)\,.
\label{sc-pot}
\ee
Qualitatively, this quantum correction to the
effective scalar potential (\ref{sc-pot}) is similar to the one in
the four-dimensional Wess-Zumino model \cite{book}.

\section{Summary}
We studied the two-loop effective action in the three-dimensional
general chiral superfield model which is described by the classical
K\"ahler potential $K(\Phi,\bar\Phi)$ and chiral potential
$W(\Phi)$. In general, this model in non-renormalizable, but
as a particular case it includes the three-dimensional Wess-Zumino
models with $K$ and $W$ given in (\ref{WZ}) which is
renormalizable. We are interested in the contributions to the
effective action up to two-derivative order which are described in
superspace by the {\it effective} K\"ahler potential $K_{\rm eff}$ and
chiral potential $W_{\rm eff}$.

At one loop order the effective K\"ahler potential receives only finite
corrections (\ref{K}) while the divergences start from two loops
(\ref{2loopresult}). Qualitatively, these two-loop results are
analogous to the one-loop K\"ahler potential in the four-dimensional
analog of the model (\ref{class}),
\cite{BP1,BP2,BCP1,BCP2,Brig,Nibbelink}.
In particular, for the three-dimensional Wess-Zumino model this effective K\"ahler
potential reduces to (\ref{K-WZ-eff}) which corresponds to the
effective scalar potential (\ref{sc-pot}). This quantum deformation of the
classical scalar potential in the three-dimensional Wess-Zumino
model is very similar to the four-dimensional case \cite{book}.

For the four-dimensional chiral
superfield model it
is well known that the effective chiral potential receives finite
quantum corrections only in the massless case
\cite{West1,West2,BKP1,BKP2,BP1,BP2,BCP1,BCP2}. We show that this
conclusion is also true for the three-dimensional chiral
superfield model under considerations. The two-loop effective
chiral potential in general chiral superfield model is given by
(\ref{WW}) and (\ref{WA}). For the particular case of the Wess-Zumino model
(\ref{WZ}), this effective chiral potential reduces to a finite
shift of the coupling constant (\ref{4.7}). It is interesting to
note that only the Feynman graph of topology $A$ at Fig.\ 1 is
responsible for this shift while in the four-dimensional case the
effective chiral potential originates form the diagram of topology
$B$ \cite{book}.

A natural extension of the present considerations might by a study
of two-loop effective action in three-dimensional $\cN=2$ supersymmetric
electrodynamics and SYM theories with matter.

\vspace{30pt}
\noindent
{\bf Acknowledgments}\\[3mm]
The authors acknowledge the support from the RFBR grant Nr.\
12-02-00121 and from LRSS grant Nr.\ 224.2012.2. I.L.B.\ and I.B.S.\
are grateful to the RFBR-Ukraine grant Nr.\ 11-02-90445 for partial
support. I.L.B.\ is grateful to CAPES for supporting his visit to
the Physics Department of Universidade Federal de Juiz de Fora where
the final part of work was done. The work of I.B.S.\ was also
supported by the Marie Curie research fellowship Nr.\ 236231,
``QuantumSupersymmetry''. I.B.S and B.S.M. acknowledge the support
by RF Federal Program ``Kadry'' under the contract 16.740.11.0469.

\appendix
\section{Details of two-loop calculations}
\subsection{Two-loop effective K\"ahler potential}
\label{A1}
Let us consider the computation of the contribution (\ref{GA})
which originates from the diagram of topology $A$ at Fig.\ 1.
It involves the fourth variational derivative of the classical
action,
\be
\frac{\delta^4 S}{\delta \Phi(z_1)\delta\Phi(z_2)\delta\bar\Phi(z_3)\delta\bar\Phi(z_4)}
=\frac14 D^2_{(3)}[K^{(4)}_{\Phi^2\bar\Phi^2}(z_3)\delta_+(z_2,z_3)]
\delta_-(z_3,z_4)\delta_+(z_1,z_2)\,.
\label{S4}
\ee
Using these delta-functions we integrate over $z_1$ and $z_2$,
\be
\Gamma_A=-\frac18\int d^5\bar z_2 d^5z_3\,
D^2_{(2)}[K^{(4)}_{\Phi^2\bar\Phi^2}(z_2)\delta_+(z_2,z_3)]
G_{+-}(z_3,z_2)G_{+-}(z_3,z_2)\,.
\ee
Now restore full superspace measure using (\ref{measure}) and
integrate over $z_3$,
\be
\Gamma_A=\frac12\int
d^7z\,K^{(4)}_{\Phi^2\bar\Phi^2}G_{+-}(z,z)G_{+-}(z,z)\,.
\ee
The propagator $G_{+-}$ is given in (\ref{prop}). At coincident
points we use the identity $D^2\bar
D^2\delta^4(\theta_1-\theta_2)|_{\theta_1=\theta_2}=16$, and compute the momentum
integral,
\bea
G_{+-}(z,z)&=&\frac1{K''_{\Phi\bar\Phi}}\frac1{\square+M^2}
\delta^3(x_1-x_2)|_{x_1=x_2}\nn\\
&=&-\frac1{K''_{\Phi\bar\Phi}}\int \frac{d^3p}{(2\pi)^2}
\frac{1}{p^2-M^2}=-\frac{i}{4\pi}\frac{|M|}{K''_{\Phi\bar\Phi}}\,,
\eea
Recall that $M^2=|W''|^2/(K''_{\Phi\bar\Phi})^2$.
As a result,
\be
\Gamma_A=-\frac1{32\pi^2}\int d^7z\frac{K^{(4)}_{\Phi^2\bar\Phi^2}|W''|^2}{
(K''_{\Phi\bar\Phi})^4}\,.
\label{GA_}
\ee

Consider the computation of the contribution (\ref{B1}) which
comes from the diagram $B$ at Fig.~1. The vertices in this diagram
are given in (\ref{S3}) and (\ref{S3-1}). We integrate four
delta-functions in these vertices and restore full superspace measure using
(\ref{measure}),
\be
\Gamma_{B_1}=-\frac12\int d^7z_2
d^5z_4\,K'''_{\Phi\bar\Phi^2}(z_2)W'''(z_4)
G_{++}(z_2,z_4)G_{-+}(z_2,z_4)G_{-+}(z_2,z_4)\,.
\label{aa}
\ee
Note that, because of $\delta_+(z_1,z_1)=-\frac14\bar D^2\delta^7(z_1-z_2)$,
the propagators $G_{+-}$ contain effectively four
Grassmann derivatives while $G_{++}$ only two. We use two Grassmann
derivatives to restore full superspace measure in (\ref{aa}) and
the remaining eight derivatives are necessary for applying the
identity (\ref{id}) twice,
\bea
\Gamma_{B_1}&=&-\frac12\int d^4\theta d^3x_1 d^3x_2
\frac{K'''_{\Phi\bar\Phi^2}W'''\bar W''}{(K''_{\Phi\bar\Phi})^4}
\left[\frac1{\square+M^2}\delta^3(x_1-x_2)\right]^3
\nn\\&=&\frac1{2(2\pi)^6}\int d^4\theta d^3p_1 d^3p_2
\frac{K'''_{\Phi\bar\Phi^2}W'''\bar W''}{(K''_{\Phi\bar\Phi})^4}
\frac1{p_1^2-M^2}\frac1{p_2^2-M^2}\frac1{(p_1+p_2)^2-M^2}
\eea
Calculating the momentum integral within the dimensional
regularization,
\be
\int \frac{d^3p_1 d^3p_2}{(2\pi)^6}\frac1{p_1^2-M^2}
\frac1{p_2^2-M^2}\frac1{(p_1+p_2)^2-M^2}=\frac{\Gamma(\epsilon)}{32\pi^2M^{2\epsilon}}
\,,
\label{2loop-int}
\ee
we get
\be
\Gamma_{B_1}=\frac1{64\pi^2}\int d^7z\frac{K'''_{\Phi\bar\Phi^2}W'''\bar W''}{
(K''_{\Phi\bar\Phi})^4}
\frac{\Gamma(\epsilon)}{M^{2\epsilon}}\,.
\label{GB1}
\ee

For computing (\ref{B2}) we need the variational derivative of the
classical action (\ref{S3}) and its conjugate. Integrating all the
delta-functions in these vertices, we get
\be
\Gamma_{B_2}=\frac16\int d^5\bar z_1 d^5z_4\,
\bar W''' W'''G^3_{-+}(z_1,z_4)
\,.
\ee
Restoring full superspace measure and applying the identity
(\ref{id}) we get
\bea
\Gamma_{B_2}&=&\frac16\int d^7z_1 d^7z_2\,
\frac{\bar W''' W'''}{(K''_{\Phi\bar\Phi})^3}
\frac1{\square+M^2}\delta^7(z_1-z_2)
\left[\frac1{\square+M^2}\frac1{16}D^2\bar
D^2\delta^7(z_1-z_2)\right]^2
\nn\\&=&\frac16\int d^4\theta d^3x_1d^3x_2\,
\frac{\bar W''' W'''}{(K''_{\Phi\bar\Phi})^3}
\left[\frac1{\square+M^2}\delta^3(x_1-x_2)\right]^3\,.
\eea
Applying the momentum integral (\ref{2loop-int}), we obtain
\bea
\Gamma_{B_2}=-\frac1{192\pi^2}\int d^7z
\frac{\bar W''' W'''}{(K''_{\Phi\bar\Phi})^3}
\frac{\Gamma(\epsilon)}{M^{2\epsilon}}\,.
\label{GB2}
\eea

The part of the effective action (\ref{B3}) involves the variational
derivative of the classical action (\ref{S3-1}) and its conjugate.
After integrating out all the delta-functions coming from the
vertices, we get
\be
\Gamma_{B_3}=-\int d^7z_1 d^7z_2\,K'''_{\Phi\bar\Phi^2}(z_1)
K'''_{\Phi^2\bar\Phi}(z_2)
G_{++}(z_1,z_2)G_{-+}(z_1,z_2)G_{--}(z_1,z_2)\,.
\ee
Taking into account the explicit form of the propagators
(\ref{prop}), we apply the identity (\ref{id}) and the momentum
integral (\ref{2loop-int}),
\bea
\Gamma_{B_3}&=&\int d^4\theta d^3x_1 d^3 x_2 \,
K'''_{\Phi\bar\Phi^2} K'''_{\Phi^2\bar\Phi}
\frac{W''\bar W''}{(K''_{\Phi\bar\Phi})^5}
\left[\frac1{\square+M^2}\delta^3(x_1-x_2) \right]^3
\nn\\&=&-\frac1{32\pi^2}\int d^7z\,
K'''_{\Phi\bar\Phi^2}K'''_{\Phi^2\bar\Phi}
\frac{W''\bar W''}{(K''_{\Phi\bar\Phi})^5}
\frac{\Gamma(\epsilon)}{M^{2\epsilon}}\,.
\label{GB3}
\eea

The expression (\ref{B4}) also involves the vertex (\ref{S3-1})
and its conjugate. After integrating out the delta-functions in
these vertices we get
\be
\Gamma_{B_4}=-\frac12\int d^7z_1 d^7z_2\,
K'''_{\Phi\bar\Phi^2}(z_1)
K'''_{\Phi^2\bar\Phi}(z_2)
G_{+-}(z_1,z_2)G_{-+}(z_1,z_2)G_{-+}(z_1,z_2)\,.
\ee
The propagators here have three $D^2$ and three $\bar D^2$
operators altogether. We can shrink down two $D^2\bar D^2$ pairs
owing to the identity (\ref{id}) and the remaining $\bar D^2 D^2$
operator generates the Dalembertian operator owing to (\ref{box}).
As a result, we have
\be
\Gamma_{B_4}=-\frac12\int d^4\theta d^3x_1 d^3 x_2
\frac{K'''_{\Phi\bar\Phi^2}K'''_{\Phi^2\bar\Phi}}{
(K''_{\Phi\bar\Phi})^3}
\frac\square{\square+M^2}\delta^3(x_1-x_2)
\left(\frac1{\square+M^2}\delta^3(x_1-x_2)\right)^2
.
\ee
The corresponding momentum integral reads
\be
\int \frac{d^3p_1d^3p_2}{(2\pi)^6}\frac{1}{p_1^2-M^2}\frac{1}{p_2^2-M^2}\frac{(p_1+p_2)^2}{(p_1+p_2)^2-M^2}
=-\frac{M^2}{16\pi^2}+\frac{M^2\,\Gamma(\epsilon)}{32\pi^2M^{2\epsilon}}\,.
\ee
As a result, we find
\be
\Gamma_{B_4}=\frac1{32\pi^2}\int d^7z
\frac{K'''_{\Phi\bar\Phi^2}K'''_{\Phi^2\bar\Phi}W''\bar W''}{
(K''_{\Phi\bar\Phi})^5}
-\frac1{64\pi^2}\int d^7z
\frac{K'''_{\Phi\bar\Phi^2}K'''_{\Phi^2\bar\Phi}W''\bar W''}{
(K''_{\Phi\bar\Phi})^5}\frac{\Gamma(\epsilon)}{M^{2\epsilon}}\,.
\label{GB4}
\ee

Finally, we point out that the expression (\ref{B5}) is complex conjugate
to (\ref{B1}),
\be
\Gamma_{B_5}=(\Gamma_{B,1})^*=\frac1{64\pi^2}\int d^7z
\frac{W''\bar W''' K'''_{\Phi^2\bar\Phi}}{(K''_{\Phi\bar\Phi})^4}
\frac{\Gamma(\epsilon)}{M^{2\epsilon}}\,.
\label{GB5}
\ee

The divergent and finite parts of these actions can be separated
with the help of the identity
\be
\frac{\Gamma(\epsilon)}{M^{2\epsilon}}=
\frac1\epsilon-\gamma-\ln M^2+O(\epsilon)\,,
\ee
where $\gamma$ is the Euler gamma constant. Summing up all
obtained expressions (\ref{GA_}), (\ref{GB1}), (\ref{GB2}),
(\ref{GB3}), (\ref{GB4}) and (\ref{GB5}), we find the two-loop
contribution to the effective K\"ahler potential,
\bea
\Gamma^{(2)}_K&=&\Gamma_{\rm div}+\Gamma_{\rm
fin}\,,\\
\Gamma_{\rm div}&=&\frac1{64\pi^2\epsilon}\int d^7z
\bigg[
\frac{K'''_{\Phi\bar\Phi^2}W'''\bar W''+K'''_{\Phi^2\bar\Phi}W''\bar W'''}{(K''_{\Phi\bar\Phi})^4}
-3\frac{K'''_{\Phi\bar\Phi^2}K'''_{\Phi^2\bar\Phi}|W''|^2}{(K''_{\Phi\bar\Phi})^5}
\bigg]\nn\\&&
-\frac1{192\pi^2\epsilon}\int d^7z\frac{|W'''|^2}{(K''_{\Phi\bar\Phi})^3}
\,,\\
\Gamma_{\rm fin}&=&\frac1{32\pi^2}\int d^7z\bigg[
-\frac{K^{(4)}_{\Phi^2\bar\Phi^2}|W''|^2}{(K''_{\Phi\bar\Phi})^4}
-\frac12(\gamma+\ln\frac{M^2}{\mu^2})
\frac{K'''_{\Phi\bar\Phi^2}W'''\bar W''+K'''_{\Phi^2\bar\Phi}W''\bar W'''}{
 (K''_{\Phi\bar\Phi})^4}
\nn\\&&
+\frac16(\gamma+\ln\frac{M^2}{\mu^2})\frac{|W'''|^2}{(K''_{\Phi\bar\Phi})^3}
+(1+\frac32\gamma+\frac32\ln\frac{M^2}{\mu^2})
\frac{K'''_{\Phi\bar\Phi^2}K'''_{\Phi^2\bar\Phi}|W''|^2}{(K''_{\Phi\bar\Phi})^5}
\bigg]\,.
\eea

\subsection{Two-loop chiral potential}
\label{A2}
\subsubsection{Diagram $a$}
Consider the computation of the two-loop Feynman graph depicted at Fig.\ 2a
with one quartic antichiral vertex $\bar\phi^4 \bar W^{(4)}$ and with
one chiral vertex $\phi^2 W''$ at each loop,
\be
\Gamma_{a}=-\frac18\int  d^5z_1 d^5 z_2d^5\bar z_3\,
W''(z_1)W''(z_2) \bar W^{(4)}(z_3)
G_{0}(z_1,z_3)G_{0}(z_1,z_3)G_{0}(z_2,z_3)G_{0}(z_2,z_3)\,,
\ee
where the propagator $G_{0}$ is given in (\ref{prop1}). From
these propagators, we use two $\bar D^2$ and one $D^2$ operators to restore full
superspace measures and put another $D^2$ on the external lines,
\bea
\Gamma_{a}&=&\frac1{32}\frac1{16^2}\int  d^7z_1 d^7 z_2d^7 z_3\,
(D^2 \frac{W''(z_1)}{(K''_{\Phi\bar\Phi}(z_1))^2})\frac{W''(z_2)}{(K''_{\Phi\bar\Phi}(z_2))^2}
 \bar W^{(4)}\nn\\&&\times
\frac1\square \delta^7(z_1-z_3)\frac{\bar D^2
D^2}{\square}\delta^7(z_1-z_3)
\frac1\square \delta^7(z_2-z_3)\frac{\bar D^2
D^2}{\square}\delta^7(z_2-z_3)\,.
\eea
We apply the identity (\ref{id}) twice and pass to the momentum
space,
\bea
\Gamma_{a}&=&\frac1{32}\int  d^4\theta \frac{d^3p d^3k_1 d^3 k_2}{(2\pi)^{9}}\,
(D^2 \frac{W''(p)}{(K''_{\Phi\bar\Phi}(p))^2})\frac{W''(-p)}{(K''_{\Phi\bar\Phi}(-p))^2}
 \bar W^{(4)}\nn\\&&\times
\frac1{k_1^2(k_1+p)^2}\frac1{k_2^2(k_2-p)^2}\,.
\eea
Computing the momentum integrals,
\be
\int \frac{d^3k}{k^2(p+k)^2}=\frac{\pi^3}{|p|}\,,
\label{simpleint}
\ee
and passing back to the coordinate space we find in the local
limit,
\be
\Gamma_{a}=-\frac1{32}\frac1{64}\int  d^7z
\frac{W'' \bar W^{(4)}}{(K''_{\Phi\bar\Phi})^2}
 \frac{D^2}{\square} \frac{W''}{(K''_{\Phi\bar\Phi})^2}\,.
\ee
Hence, the corresponding contribution to the effective chiral
potential is
\be
\Gamma_{a}=\frac1{512}\int  d^5z
\frac{ \bar W^{(4)} (W'')^2}{(K''_{\Phi\bar\Phi})^4}\,.
\ee
Note that $\bar W^{(4)}$ is constant here and $K''_{\Phi\bar\Phi}$
is considered at $\bar\Phi=0$ and therefore it is chiral.

\subsubsection{Diagram $b_1$}

In the diagram $b_1$ the vertices ``1'' and ``3'' are chiral, ``4'' is the
antichiral vertex and ``2'' is the vertex in the full superspace.
Therefore, this diagram corresponds to the following expression
\bea
\Gamma_{b_1}&=&\frac12\int d^5z_1 d^7z_2 d^5 z_3 d^5\bar z_4\,
K'''_{\bar\Phi^2\Phi}(z_2) W''(z_1) W''(z_3) \bar W'''
\nn\\&&\times
G_{0}(z_1,z_2)G_{0}(z_1,z_4)G_{0}(z_3,z_2)G_{0}(z_3,z_4)G_{0}(z_2,z_4)\,.
\eea
We use one $D^2$ and two $\bar D^2$ operators from the propagators
to restore the full superspace measure and integrate by parts another two
$D^2$ operators,
\bea
\Gamma_{b_1}&=&-\frac{1}{8\cdot 16^3}\int d^7z_1\ldots d^7z_4\,
K'''_{\bar\Phi^2\Phi}(z_2)\bar W'''
\frac1\square\delta^7(z_1-z_2)\frac1\square\delta^7(z_3-z_2)
\frac{\bar D^2}{K''_{\Phi\bar\Phi}\square}\delta^7(z_2-z_4)
\nn\\&&\times
D^2\left(
\frac{W''(z_1)}{(K''_{\Phi\bar\Phi})^2}\frac{\bar D^2
D^2}{\square}\delta^7(z_1-z_4)
\right)
D^2\left(
\frac{W''(z_3)}{(K''_{\Phi\bar\Phi})^2}
\frac{\bar D^2 D^2}{\square}\delta^7(z_3-z_4)
\right)\,.
\label{67}
\eea
Using the delta-functions we integrate over the Grassmann
variables $\theta_1$ and $\theta_3$ and distribute the
$D$-operators in the second line in (\ref{67}) keeping only the
terms containing no more than two derivatives $D_\alpha$ on the
external lines,
\bea
\Gamma_{b_1}&=&-\frac{1}{8\cdot 16^2}\int d^3x_1 d^7z_2 d^3x_3 d^7z_4\,
K'''_{\bar\Phi^2\Phi}(z_2)\bar W'''
\frac1\square\delta^3(x_1-x_2)\frac1\square\delta^3(x_3-x_2)
\nn\\&&\times
\frac{\bar D^2}{K''_{\Phi\bar\Phi}\square}\delta^7(z_2-z_4)
[T_1+T_2+T_2]\,,
\label{68}
\eea
where
\bea
T_1&=&
D^2\left(\frac{W''(x_1,\theta_2)}{(K''_{\Phi\bar\Phi})^2}\right)
\frac{\bar D^2 D^2}\square
\left(\delta^3(x_1-x_4)\delta^4(\theta_2-\theta_4)\right)
\nn\\&&\times
\frac{W''(x_3,\theta_2)}{(K''_{\Phi\bar\Phi})^2}D^2
\left(\delta^3(x_3-x_4)\delta^4(\theta_2-\theta_4)\right)\,,
\nn\\
T_2&=&
\frac{W''(x_1,\theta_2)}{(K''_{\Phi\bar\Phi})^2}
 D^2\left( \delta^3(x_1-x_4)\delta^4(\theta_2-\theta_4)\right)
\nn\\&&\times
D^2\left(\frac{W''(x_3,\theta_2)}{(K''_{\Phi\bar\Phi})^2}\right)
\frac{\bar D^2 D^2}\square\left(
\delta^3(x_3-x_4)\delta^4(\theta_2-\theta_4)\right)\,,
\nn\\
T_3&=&
-4D^\alpha \left(\frac{W''(x_1,\theta_2)}{(K''_{\Phi\bar\Phi})^2}\right)
\frac{\partial_{\alpha\beta} \bar D^\beta D^2}{\square}
\delta^3(x_1-x_2)\delta^4(\theta_2-\theta_4)
\nn\\&&\times
D^\gamma\left(\frac{W''(x_3,\theta_2)}{(K''_{\Phi\bar\Phi})^2}\right)
\frac{\partial_{\gamma\delta}\bar D^\delta D^2}{\square}
\delta^3(x_3-x_4)\delta^4(\theta_2-\theta_4)\,.
\eea
We integrate by parts the operator $\bar D^2$ in the first line of
(\ref{68}) and apply the identity (\ref{id}) twice. Then we take
another $\bar D^2$ from the full measure which can hit only the
external lines and pass to the momentum space,
\be
\Gamma_{b_1}=-\frac12\int \frac{d^3p_1 d^3p_2 d^2\theta}{(2\pi)^6}
\frac{W''(p_1,\theta)W''(p_2,\theta)K'''_{\bar\Phi^2\Phi}(-p_1-p_2,\theta)
\bar W'''}{(K''_{\Phi\bar\Phi})^5}
S_1(p_1,p_2)\,,
\label{69}
\ee
where
\be
S_1(p_1,p_2)=\int \frac{d^3k_1 d^3k_2}{(2\pi)^6}
\frac{p_1^2 k_2^2+k_1^2 p_2^2-2(p_1p_2)(k_1 k_2)
+2(p_1 k_2)(p_2 k_1)-2(p_1k_1)(p_2k_2)}{k_1^2 k_2^2(p_1+k_1)^2
(p_2+k_2)^2(k_1+k_2)^2}\,.
\label{70}
\ee
We are interested in the local contributions to the effective
action. To take the local limit we make the inverse Fourier transform for the
superfields,
\bea
\Gamma_{b_1}&=&-\frac12\int d^2\theta d^3x_1 d^3x_2 d^3x_3
\frac{W''(x_1,\theta)W''(x_2,\theta)K'''_{\bar\Phi^2\Phi}(x_3,\theta)
\bar W'''}{(K''_{\Phi\bar\Phi})^5}\nn\\
&&\times \int \frac{d^3p_1 d^3p_2}{(2\pi)^6}
S_1(p_1,p_2)e^{ix_1 p_1 }e^{ix_2 p_2} e^{-ix_3(p_1+p_2)}\,,
\label{69.1}
\eea
and assume that the superfields vary slowly in the Minkowski space,
\be
W''(x_1,\theta)W''(x_2,\theta)K'''_{\bar\Phi^2\Phi}(x_3,\theta)
\simeq
W''(x_1,\theta)W''(x_1,\theta)K'''_{\bar\Phi^2\Phi}(x_1,\theta)\,.
\ee
The integration over $d^3x_2$ and $d^3x_3$ in (\ref{69.1}) yields
two delta-functions, $(2\pi)^3\delta^3(p_2)$ and
$(2\pi)^3\delta^3(p_1+p_2)$. These delta-functions show that we
need to compute the limit
\be
S_1=\lim_{p_1\to0,p_2\to 0}S_1(p_1,p_2)\,.
\ee
However, in the Appendix \ref{appB2} we show that this limit does
not exist, i.e., the value of $S_1$ depends essentially on the way
of computing this limit. Hence, we conclude that the diagram $b_1$
in Fig.~2 does not give any local contribution to the effective
action in the chiral sector.

\subsubsection{Diagram $b_2$}
\label{app-b2}
This diagram has the chiral vertices ``1'', ``3'' and the antichiral
ones ``4'', ``2''. The
corresponding analytic expression reads
\bea
\Gamma_{b_2}&=&-\frac12\int d^5z_1 d^5\bar z_2 d^5 z_3 d^5\bar z_4\,
W'''(z_1)(\frac14D^2K''_{\bar\Phi^2}(z_2))W''(z_3)\bar W'''
\nn\\&&\times
G_{0}(z_1,z_2)G_{0}(z_3,z_2)G_{0}(z_3,z_4)G_{0}(z_1,z_4)
G_{0}(z_1,z_4)\,.
\eea
Restore the full superspace measures using the $D^2$-operators
from the propagators,
\bea
\Gamma_{b_2}&=&-\frac1{2\cdot 16^3}\int d^7z_1 d^7z_2 d^7 z_3 d^7 z_4\,
W'''(z_1)(\frac14D^2 K''_{\bar\Phi^2}(z_2))W''(z_3)\bar W'''
\frac{1}{K''_{\Phi\bar\Phi}\square}\delta^7(z_1-z_2)
\nn\\&&\times
\frac{\bar D^2 D^2}{K''_{\Phi\bar\Phi}\square}\delta^7(z_3-z_2)
\frac{1}{K''_{\Phi\bar\Phi}\square}\delta^7(z_3-z_4)
\frac{\bar D^2 D^2}{K''_{\Phi\bar\Phi}\square}\delta^7(z_1-z_4)
\frac{\bar D^2 D^2}{K''_{\Phi\bar\Phi}\square}\delta^7(z_1-z_4)
\,.
\label{74}
\eea
Note that in the first line of this expression we have the
operator $D^2$ on the external line $K''_{\bar\Phi^2}$ and, hence,
all other terms with derivatives on the external lines
can be neglected. This means that upon integration by parts the
Grassmann derivatives do not hit the external lines. The
operator $\bar D^2 D^2$ on the delta-function $\delta^7(z_3-z_2)$ in the second line
of (\ref{74}) can be transported to the delta functions $\delta^7(z_1-z_4)$
and then it produces the box operator owing to the
identity (\ref{box}),
\bea
\Gamma_{b_2}&=&-\frac1{2\cdot 16^2}\int d^7z_1 d^7z_2 d^7 z_3 d^7 z_4\,
W'''(z_1)(\frac14D^2 K''_{\bar\Phi^2}(z_2))W''(z_3)\bar W'''
\frac{1}{(K''_{\Phi\bar\Phi})^5}
\label{75}
\\&&\times\frac1{\square}\delta^7(z_1-z_2)
\frac{1}{\square}\delta^7(z_3-z_2)
\frac{1}{\square}\delta^7(z_3-z_4)
\square\left(
\frac{\bar D^2 D^2}{\square}\delta^7(z_1-z_4)
\frac{\bar D^2
D^2}{\square}\delta^7(z_1-z_4)\right)
.\nn
\eea
We integrate over all but one Grassmann variables and apply the
identity (\ref{id}) twice,
\bea
\Gamma_{b_2}&=&-\frac1{2}\int d^4\theta d^3x_1 d^3x_2 d^3 x_3 d^3 x_4\,
\frac{W'''(x_1,\theta)(\frac14D^2 K''_{\bar\Phi^2}(x_2,\theta))W''(x_3,\theta)\bar
W'''}{(K''_{\Phi\bar\Phi})^5}
\label{76}\\&&\times
\frac{1}{\square}\delta^3(x_1-x_2)
\frac{1}{\square}\delta^3(x_2-x_3)
\frac{1}{\square}\delta^3(x_3-x_4)
\square\left(\frac{1}{\square}\delta^3(x_1-x_4)
\frac{1}{\square}\delta^3(x_1-x_4)\right)
\,.\nn
\eea
In this expression we pass to the chiral subspace for the
Grassmann variables and to the momentum representation for the
Minkowski space coordinates,
\be
\Gamma_{b_2}=-\frac1{2}\int d^2\theta \frac{d^3p_1 d^3p_2}{(2\pi)^6}\,
\frac{W'''(-p_1-p_2,\theta) K''_{\bar\Phi^2}(p_1,\theta)W''(p_2,\theta)\bar
W'''}{(K''_{\Phi\bar\Phi})^5}
S_2(p_1,p_2)\,,
\label{77}
\ee
where
\be
S_2(p_1,p_2)=\int\frac{d^3k_1 d^3k_2}{(2\pi)^6}\frac{p_1^2}{(p_1+p_2+k_1)^2
(p_2+k_1)^2(k_1+k_2)^2 k_2^2}\,.
\label{78}
\ee
In this function, the integration over $k_2$ can be done using
(\ref{simpleint}),
\be
S_2(p_1,p_2)=\int \frac{d^3k}{(4\pi)^3}\frac{p_1^2}{
(p_1+p_2+k)^2(p_2+k)^2|k|}\,.
\label{79}
\ee
In principle, this momentum integral can be computed for general
values of the external momenta, but we need to find the local limit
to single out the contributions to the effective chiral potential.
We make the inverse Fourier transform for the superfields in (\ref{77}),
\bea
\Gamma_{b_2}&=&-\frac1{2}\int d^2\theta d^3x_1 d^3x_2 d^3x_3\,
\frac{W'''(x_1,\theta) K''_{\bar\Phi^2}(x_2,\theta)W''(x_3,\theta)\bar
W'''}{(K''_{\Phi\bar\Phi})^5}
\nn\\&&\times
\int\frac{d^3p_1 d^3p_2}{(2\pi)^6} e^{-ix_1(p_1+p_2)}e^{ix_2p_1}e^{ix_3 p_2}
S_2(p_1,p_2)\,,
\label{77.1}
\eea
and consider the slowly varying fields,
\be
W'''(x_1,\theta) K''_{\bar\Phi^2}(x_2,\theta)W''(x_3,\theta)
\simeq W'''(x_3,\theta) K''_{\bar\Phi^2}(x_3,\theta)W''(x_3,\theta)\,.
\ee
The integration over $d^3x_1$ and $d^3x_2$ yields two delta
functions, $(2\pi)^3\delta^3(p_1+p_2)$ and
$(2\pi)^3\delta^3(p_2)$. These delta-functions mean that we need
to compute the limit
\be
\lim_{p_1\to 0,p_2\to0}S_2(p_1,p_2)\,.
\label{lim}
\ee
In the Appendix \ref{int79} we show that this limit does not
exist, i.e., its value depend on the way of computing this limit.
Hence, we conclude that the Feynman graph $b_2$ in Fig.~2 does con
contribute to the effective chiral potential.

\subsubsection{Diagram $b_3$}
This diagram has the vertex ``1'' in the full superspace, the
vertices ``2'' and ``4'' in the chiral subspace while ``3'' and
``5'' are antichiral vertices. The corresponding expression reads
\bea
\Gamma_{b_3}&=&\frac12 \int d^7z_1 d^5z_2 d^5\bar z_3 d^5z_4
d^5\bar z_5\, K'''_{\Phi^2\bar\Phi}(z_1)W''(z_2)
(\frac14D^2 K''_{\bar\Phi^2}(z_3))W''(z_4)\bar W'''
\nn\\&&\times
G_0(z_2,z_1)G_{0}(z_2,z_3)G_0(z_4,z_3)G_{0}(z_4,z_5)
G_{0}(z_1,z_5)G_{0}(z_1,z_5)\,.
\label{80}
\eea
As the first step, we restore the full superspace measure in
(\ref{80}) by taking the Grassmann derivatives from the
propagators,
\bea
\Gamma_{b_3}&=&\frac12\frac1{16^4} \int d^7z_1 d^7z_2 d^7 z_3 d^7z_4
d^7 z_5\, K'''_{\Phi^2\bar\Phi}(z_1)W''(z_2)
(\frac14D^2 K''_{\bar\Phi^2}(z_3))W''(z_4)\bar W'''
\nn\\&&\times
\frac{\bar D^2 D^2}{K''_{\Phi\bar\Phi}\square}\delta^7(z_2-z_1)
\frac{1}{K''_{\Phi\bar\Phi}\square}\delta^7(z_2-z_3)
\frac{\bar D^2 D^2}{K''_{\Phi\bar\Phi}\square}\delta^7(z_4-z_3)
\frac{1}{K''_{\Phi\bar\Phi}\square}\delta^7(z_4-z_5)
\nn\\&&\times
\frac{\bar D^2 D^2}{K''_{\Phi\bar\Phi}\square}\delta^7(z_1-z_5)
\frac{\bar D^2 D^2}{K''_{\Phi\bar\Phi}\square}\delta^7(z_1-z_5)
\,.
\label{81}
\eea
In the first line here we have the expression $\frac14 D^2
K''_{\bar\Phi^2}$ which shows that we have sufficient number of
Grassmann derivatives on the external lines to apply the identity
(\ref{no-nonren}). Hence, upon integration by parts of the
other Grassmann derivatives we can omit all the terms in which these
derivatives hit the external lines.
For instance, in the second line of (\ref{81}) we integrate by parts
the derivatives $\bar D^2 D^2$ acting on $\delta^7(z_4-z_3)$
such that they hit $\delta^7(z_2-z_1)$ and
cancel one box operator,
\bea
\Gamma_{b_3}&=&\frac12\frac1{16^3} \int d^7z_1 d^7z_2 d^7 z_3 d^7z_4
d^7 z_5\, K'''_{\Phi^2\bar\Phi}(z_1)W''(z_2)
(\frac14D^2 K''_{\bar\Phi^2}(z_3))W''(z_4)\bar W'''
\frac1{(K''_{\Phi\bar\Phi})^6}
\nn\\&&\times
\bar D^2 D^2\delta^7(z_2-z_1)
\frac{1}{\square}\delta^7(z_2-z_3)
\frac{1}{\square}\delta^7(z_4-z_3)
\frac{1}{\square}\delta^7(z_4-z_5)
\nn\\&&\times
\frac{\bar D^2 D^2}{\square}\delta^7(z_1-z_5)
\frac{\bar D^2 D^2}{\square}\delta^7(z_1-z_5)
\,.
\label{82}
\eea
In a similar way we integrate by parts the derivatives $\bar D^2
D^2$ in the second line of (\ref{82}) such that they hit the delta
functions $\delta^7(z_1-z_5)$ and yield one more box operator,
\bea
\Gamma_{b_3}&=&\frac12\frac1{16^2} \int d^7z_1 d^7 z_2 d^7z_3
d^7 z_4\, K'''_{\Phi^2\bar\Phi}(z_1)W''(z_1)
(\frac14D^2 K''_{\bar\Phi^2}(z_2))W''(z_3)\bar W'''
\frac1{(K''_{\Phi\bar\Phi})^6}
\nn\\&&\times
\frac{1}{\square}\delta^7(z_1-z_2)
\frac{1}{\square}\delta^7(z_3-z_2)
\frac{1}{\square}\delta^7(z_3-z_4)
\square\left(
\frac{\bar D^2 D^2}{\square}\delta^7(z_1-z_4)
\frac{\bar D^2 D^2}{\square}\delta^7(z_1-z_4)\right)
\,.
\nn\\
\label{82-1}
\eea
Now we apply the identity (\ref{id}) two times, integrate all but
one Grassmann variables and pass to the momentum space for the
Minkowski space coordinates,
\bea
\Gamma_{b_3}&=&\frac12 \int d^2\theta \frac{d^3p_1 d^3p_2}{(2\pi)^6}
\, \frac{K'''_{\Phi^2\bar\Phi}(-p_1-p_2,\theta)W''(-p_1-p_2,\theta)
K''_{\bar\Phi^2}(p_1,\theta)W''(p_2,\theta)\bar W'''}{(K''_{\Phi\bar\Phi})^6}
\nn\\&&\times
S_2(p_1,p_2)\,,
\eea
where the function $S_2(p_1,p_2)$ is given by the momentum
integral (\ref{79}). As is proved in the Appendix \ref{int79},
this function does not have the local limit, (\ref{lim1}).
Hence, we conclude that the diagram $b_3$ does not contribute to
the effective chiral potential.

\section{Momentum integrals}
\subsection{Local limit for the momentum integral (\ref{78})}
\label{int79}

It is hard to compute the momentum integral (\ref{78}) for
arbitrary values of external momenta $p_1$ and $p_2$. In fact, we
need only the local limit (\ref{lim}) for this integral. Our aim
is to prove that this function $S_2(p_1,p_2)$ does not possess
well defined local limit and, hence, does not contribute to the
effective chiral potential.

We are going to demonstrate that the limit (\ref{lim}) depends
essentially on the path tending to the origin of the $(p_1,p_2)$
space. Hence, choosing different ways of computing this limit, the
expression (\ref{lim}) can be given an arbitrary value, including
zero and infinity. In fact, it is sufficient to compute (\ref{78})
for $p_2=tp_1\equiv tp$, with some real parameter $t$, and to show
that the momentum integral (\ref{78}) acquires different values
for different $t$, with arbitrary small $p$.

Let us represent (\ref{78}) as
\be
S_2(p_1,p_2)=p_1^2 S(p_1,p_2)\,,
\ee
where
\be
S(p_1,p_2)=\int\frac{d^3k_1 d^3k_2}{(2\pi)^6}\frac{1}{(p_1+p_2+k_1)^2
(p_2+k_1)^2(k_1+k_2)^2 k_2^2}\,.
\label{B_1}
\ee
Using (\ref{simpleint}) we integrate over $k_2$,
\be
S(p_1,p_2)=\int \frac{d^3k}{(4\pi)^3}\frac{1}{
(p_1+p_2+k)^2(p_2+k)^2|k|}\,.
\label{B_2}
\ee
Introducing the Feynman parameters we merge together the
propagators,
\bea
S(p_1,p_2)&=&\frac34\int \frac{d^3k}{(4\pi)^3}
\int_0^1\frac{d\alpha}{\sqrt\alpha} \int_0^{1-\alpha}
d\beta
\frac{1}{
[\alpha k^2+\beta(p_1+p_2+k)^2 +(1-\alpha-\beta)(p_2+k)^2]^{5/2}}
\nn\\&=&\frac34
\int_0^1\frac{d\alpha}{\sqrt\alpha} \int_0^{1-\alpha}
d\beta
\int \frac{d^3k}{(4\pi)^3}
\frac{1}{
P^{5/2}}\,,
\eea
where
\be
P=k^2+2k(\beta(p_1+p_2)+(1-\alpha-\beta)p_2)
+\beta(p_1+p_2)^2+(1-\alpha-\beta)p_2^2\,.
\ee

Using the general formula \cite{Frampton}
\be
\int \frac{d^3k}{(k^2+2k\cdot Q -M^2)^{\alpha}}
=\frac{i\pi^{3/2}(-1)^{-\alpha}\Gamma(\alpha-3/2)}{\Gamma(\alpha)
(M^2+Q^2)^{\alpha-3/2}}\,,
\label{Fr}
\ee
we compute the integral over momenta,
\be
S(p_1,p_2)=\frac1{64\pi^2}
\int_0^1\frac{d\alpha}{\sqrt\alpha} \int_0^{1-\alpha}
d\beta
\frac{1}{\beta(\beta-1)p_1^2+\alpha(\alpha-1)p_2^2
-2\alpha\beta (p_1\cdot p_2)
}\,.
\ee
The integration over $\beta$ can de done readily,
\be
S(p_1,p_2)=\frac1{64\pi^2p_1^2}
\int_0^1\frac{d\alpha}{\sqrt\alpha}
\frac{1}{2M}
\ln\left|
\frac{M-1/2+\alpha Y+\alpha}{M+1/2-\alpha Y-\alpha}
\frac{M-1/2-\alpha Y}{M+1/2+\alpha Y}
\right|
\,,
\label{B_5}
\ee
where
\be
X=\frac{p_2^2}{p_1^2}\,,\quad
Y=\frac{(p_1\cdot p_2)}{p_1^2}\,,\quad
M=\sqrt{(1/2+\alpha Y)^2+\alpha(1-\alpha)X}.
\ee

The integral over the remaining parameter $\alpha$ is very
complicated. To simplify this, we consider the collinear momenta,
$p_2=tp_1\equiv tp$. In this case $X=t^2$, $Y=t$, $M^2=\frac14+\alpha
t(t+1)$, and the integral (\ref{B_5}) simplifies,
\be
S_2(p,tp)=\frac1{32\pi^2\sqrt{t(t+1)}}\int_{1/2}^{t+1/2}
\frac{dM}{\sqrt{M-1/2}\sqrt{M+1/2}}
\ln\frac{M-1/2}{M+1/2}\,.
\ee
The value of this integral can be expressed in terms of the Euler
di-logarithm function,
\be
S_2(t)\equiv S_2(p,tp)=
\frac1{16\pi^2\sqrt{t(t+1)}}
\left[
{\rm Li}_2\frac{\sqrt{1+t}-\sqrt t}{\sqrt{1+t}+\sqrt t}
-{\rm Li}_2\left(-\frac{\sqrt{1+t}-\sqrt t}{\sqrt{1+t}+\sqrt
t}\right)-\frac{\pi^2}{4}
\right]
\,.
\label{func}
\ee
The
function $S_2(t)$ is well defined for $0<t<\infty$.
Here we see
that all momenta cancelled out and the remaining function
depends only on the parameter $t$ which relates the momenta.
Hence, the value of $S_2(p,tp)$ remains the same for arbitrary small momenta,
$p\to 0$, but depends essentially on $t$.
As a result, along the path $p_2=tp_1$ the limit
(\ref{lim}) depends essentially on the parameter $t$ and
it can take arbitrary value allowed for the function (\ref{func}).
This proves that
\be
\lim_{p_1\to 0,p_2\to0}S_2(p_1,p_2) \quad\mbox{does not exist}\,.
\label{lim1}
\ee

\subsection{Local limit for the momentum integral (\ref{70})}
\label{appB2}
Similarly as in the previous subsection we are going to prove that
\be
\lim_{p_1\to 0,p_2\to0}S_1(p_1,p_2) \quad\mbox{does not exist}\,.
\label{lim2}
\ee
To show this it is sufficient to compute (\ref{70}) for collinear
momenta, $p_1=tp_2\equiv tp$ for which it simplifies,
\be
S_1(tp,p)=p^2\int \frac{d^3k_1 d^3 k_2}{(2\pi)^6}
\frac{t^2 k_2^2+k_1^2-2t(k_1k_2)}{k_1^2 k_2^2 (tp+k_1)^2(p+k_2)^2(k_1+k_2)^2}
=I_1+I_2+I_3\,,
\label{B_14}
\ee
where
\bea
I_1&=&p^2t(t+1)\int \frac{d^3k_1 d^3 k_2}{(2\pi)^6}
\frac{1}{k_1^2 (tp+k_1)^2(p+k_2)^2(k_1+k_2)^2}\,,\\
I_2&=&p^2(t+1)\int \frac{d^3k_1 d^3 k_2}{(2\pi)^6}
\frac{1}{k_2^2 (tp+k_1)^2(p+k_2)^2(k_1+k_2)^2}\,,\\
I_3&=&-tp^2\int \frac{d^3k_1 d^3 k_2}{(2\pi)^6}
\frac{1}{k_1^2 k_2^2(tp+k_1)^2(p+k_2)^2}\,.
\eea
It is easy to see that the last integral is just a constant,
\be
I_3=-\frac1{64}\,,
\label{I_3}
\ee
where (\ref{simpleint}) has been used. For the remaining two
integrals, the integration over one of the momenta can be done readily,
\bea
I_1(t)&=&p^2t(t+1)\int \frac{d^3k}{(4\pi)^3}
\frac{1}{|k| (p+k)^2(k+p(t+1))^2}\,,\\
I_2(t)&=&p^2(t+1)\int \frac{d^3k}{(4\pi)^3}
\frac{1}{|k|(tp+k)^2(k+p(t+1))^2}=I_1(1/t)\,.
\eea
This integrals can be calculated using (\ref{B_5}) and
(\ref{func}),
\bea
I_1&=&\frac{\sqrt{t(t+1)}}{16\pi^2}
\left[
{\rm Li}_2\frac{\sqrt{1+t}-\sqrt t}{\sqrt{1+t}+\sqrt t}
-{\rm Li}_2\left(-\frac{\sqrt{1+t}-\sqrt t}{\sqrt{1+t}+\sqrt
t}\right)-\frac{\pi^2}{4}
\right]
\label{I_1}
\,,\\
I_2&=&\frac{\sqrt{t+1}}{16\pi^2 t}
\left[
{\rm Li}_2\frac{\sqrt{1+t}-1}{\sqrt{1+t}+1}
-{\rm Li}_2\left(-\frac{\sqrt{1+t}-1}{\sqrt{1+t}+1}\right)
-\frac{\pi^2}{4} \right]
\,.
\label{I_2}
\eea

As a result, the value of the integral (\ref{B_14}) is given by
the sum of (\ref{I_3}), (\ref{I_1}) and (\ref{I_2}). It depends
only on the parameter $t$ rather than on the momentum $p$. Hence,
its value remains the same for arbitrary small momentum $p$, but
it changes upon varying the parameter $t$. This proves
(\ref{lim2}).

\end{document}